\documentclass[aps,prx,superscriptaddress,
longbibliography,
twocolumn]{revtex4-1}

\usepackage{amsfonts}
\usepackage{subfigure}
\usepackage{amsmath}
\usepackage{txfonts}
\usepackage{amssymb}
\usepackage{amsbsy} 
\usepackage{epsfig}
\usepackage{graphicx}
\usepackage{float}
\usepackage{epstopdf}

\def\Re{\rm{Re}}
\def\Im{\rm{Im}}
\def\be{\begin{equation}} \def\ee{\end{equation}}
\def\bea{\begin{eqnarray}} \def\eea{\end{eqnarray}}

\def\bk{{\bf k}}

\def\bB{{\bf B}}

\def\be{{\bf e}}

\def\bA{{\bf A}}

\def\rw{\rightarrow}

\begin{document}

\title{Magneto-optical conductivity of double-Weyl semimetals}

\author{Yong Sun}
\affiliation{ Institute for Theoretical Physics and Department of Modern Physics
University of Science and Technology of China, Hefei, 230026, P. R. China}
\altaffiliation{ suninsky@mail.ustc.edu.cn}

\author{An-Min Wang}
\affiliation{ Institute for Theoretical Physics and Department of Modern Physics
University of Science and Technology of China, Hefei, 230026, P. R. China}


\begin{abstract}
We have investigated the magneto-optical response of double Weyl semimetals whose energy
dispersion is intrinsically anisotropic. We find that in the presence of a magnetic field,
the most salient feature of the optical conductivity is a series of
resonant peaks with the corresponding frequencies scaling linearly with the strength of the magnetic field. Besides,
the optical conductivity is also found to be anisotropic, with two of the three longitudinal components
residing at a linear background and the rest one at a constant background.
The effects of chemical potential, temperature, impurity scattering and particle-hole symmetry
breaking to
the optical conductivity are also studied.
\end{abstract}

\maketitle

\section{Introduction}
Recently, topological semimetals have been actively
studied because of the existence of nontrivial band crossings
protected by topology and symmetries~\cite{Chiu2015RMP}. According to the dimension
and degeneracy of the band crossings, the currently most studied topological
semimetals can be classified into three classes: Weyl semimetals~\cite{Murakami2007weyl,wan2011,burkov2011,
weng2015,Huang2015TaAs,Xu2015weyl,lv2015,lu2015}, Dirac semimetals~\cite{young2012dirac,wang2012dirac,wang2013three,neupane2014,xu2015observation,liu2014discovery,Borisenko2014}
and nodal line semimetals~\cite{Burkov2011nodal,Bzdusek2016,
hopflink, nodal-link, nodal-knot}. For the Weyl semimetals, the band crossings are
isolated points (zero dimension) with two-fold degeneracy, known as the Weyl points,
and play the role of monopoles in the Brillouin zone. As the net charge of
the monopoles in the Brillouin zone must be zero, the number of Weyl points
with opposite monopole charge must be equal~\cite{NIELSEN1981}.

The monopole charge of a Weyl point is equal to the number of Berry fluxes
passing through a closed surface enclosing the target Weyl
point only. So far, most studies on Weyl semimetals have focused on
the case with the lowest monopole charge, $C=\pm1$ (we will refer to
Weyl semimetals with $C=\pm1$ as single Weyl semimetals for
notational simplicity)~\cite{Hosur2013,Yan2016review, Lu2016review, Hasan2017review, Burkov2017review}. The reason that
this case has attracted special research interest is because the energy dispersion
away from the Weyl points is linear in all directions and thus an analog
of the elementary Weyl fermions in particle physics, furthermore, it is the one
found in condensed matter experiments to date~\cite{Hasan2017review}.
However, the much more unexplored Weyl points with higher monopole
charge~\cite{Xu2011doubleweyl,Fang2012MW, Huang2016double,Lai2015dwsm,Jian2015dwsm,Zhang2016triple,
Wang2016triple,Chen2016doubleweyl,Ahn2016multiWeyla,Ahn2017mw,Hayata2017MW,Gupta2017MW,roy_magnetic_2015,li_weyl_2016} are
also of fundamental interest in  at least two aspects. First, the increase of monopole
charge no doubt will enhance a series of effects predicted or observed
in single Weyl semimetals, like the anomalous Hall effect~\cite{Yang2011,Burkov2014AHE,Chan2016hall,Yan2016tunable},
and the ones related to chiral anomaly~\cite{son2012,liu2012,aji2011,Zyuzin2012anomaly,wang2013a,Hosur-anomaly,
Kim-chiral-anomaly,Parameswaran-anomaly}.
Second, the energy dispersion will become intrinsically anisotropic
for higher monopole charge, which consequently will induce novel physics absent
in single Weyl semimetals. For instance, it is found that Weyl points with higher
monopole charge exhibit
anisotropic screening to the Coulomb interactions~\cite{Lai2015dwsm,Jian2015dwsm,Zhang2016triple,Wang2016triple}.

In this work, we will investigate the magneto-optical conductivity of
Weyl semimetals with $C=\pm2$, the so-called double Weyl semimetals, of
which HgCr$_{2}$Se$_{4}$~\cite{Xu2011doubleweyl,Fang2012MW} and SrSi$_{2}$~\cite{Huang2016double} were predicted as candidate materials.
In the presence of a magnetic field, the continuum spectrum of the Weyl Hamiltonian
will transform to a series of discrete Landau levels whose extrema have
divergent density of states. When optical transition between
two Landau levels takes place at their extrema, a resonant peak
will show up in the real part of the optical conductivity. In experiments,
from these resonant peaks, information, like the energy gap and the Fermi velocity,
of the underlying band structure can be extracted~\cite{Jiang2007moc,Deacon2007moc,
Plochocka2008moc,Schafgans2012moc,Orlita2015MOC,Chen2016ZrTe5,Akrap2016MOC}.
For instance, it is known
that when the frequencies of the resonant
peaks ($\omega$) and the magnetic field ($B$) follow an $\omega\propto\sqrt{B}$ law, the
energy dispersion of the Hamiltonian is gapless and linear, like those in the Dirac semimetals
and single Weyl semimetals~\cite{Gusynin2007MOC, Ashby2013weyl, Malcolm2014moc}.
For the double Weyl semimetals concerned in this paper, we find the resonant peaks of the optical conductivity
exhibit several distinctive features which can thus be applied to determine whether a material
is double Weyl semimetal or not in experiments.

The paper is organized as follows. In Sec.~\ref{Formalism}, we give
the Hamiltonian and the formula of the optical conductivity. In Sec.~\ref{MOC},
numerical results of the conductivity are given. In Sec.~\ref{ee}, we give our
estimation of the conditions needed for the magneto-optical experiments. Discussions and
conclusions are presented in Sec.~\ref{dc}.

\section{Formalism}
\label{Formalism}

As Weyl points are crossings of two bands without degeneracy, the low-energy
effective Hamiltonian of double-Weyl semimetals is given by ($\hbar=c=k_{B}=1$)
\begin{eqnarray}
H_{\chi}(\bk)=\lambda(k_{x}^{2}-k_{y}^{2})\tau_{x}+2\lambda k_{x}k_{y}\tau_{y}+\chi vk_{z}\tau_{z},\label{dw}
\end{eqnarray}
where $\tau_{i}$ are Pauli matrices, $\lambda$ is a constant parameter with the dimension of inverse mass,
$v$ refers to the velocity in the direction of linear dispersion and
$k_{x,y,z}$ refer to the momenta relative to the Weyl points;
$\chi=\pm1$ denote two kinds of chirality.
Without loss of generality, $v$ and $\lambda$ are assumed to be positive.
The energy spectra can be readily obtained,
\begin{eqnarray}
E_{\pm,\chi}(\bk)=\pm\sqrt{v^{2}k_{z}^{2}+\lambda^{2}(k_{x}^{2}+k_{y}^{2})^{2}},
\end{eqnarray}
which are linear along the $z$ direction and quadratic in the $x$-$y$ plane.

\begin{figure}
\includegraphics[width=8cm, height=5cm]{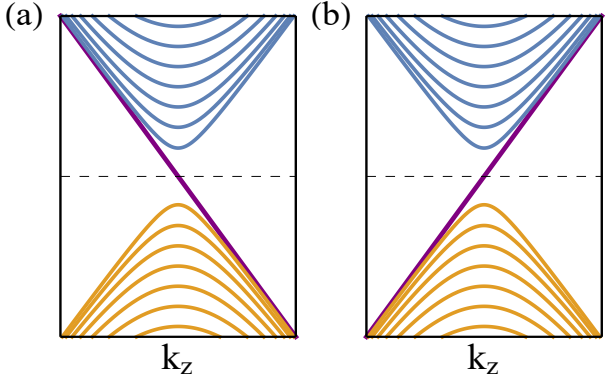}
\caption{The Landau levels of double-Weyl semimetals. (a) $\chi=1$; (b) $\chi=-1$. The Landau levels
in  blue (yellow) color belongs to the particle (hole) branch. For
the particle (hole) branch, the Landau level index $n_{p}$ ($n_{h}$) is $2$, $3$,... from bottom (top) to top (bottom). In both figures,
the chiral Landau levels ($n=0, 1$) are in purple color and doubly degenerate. The dashed
line represents the Fermi level.} \label{LL}
\end{figure}

Now we consider a magnetic field that is exerted along the $z$ direction, $\bB=B\hat{z}$.
Following the standard minimal coupling, i.e., $\bk\rw\mathbf{\Pi}=\bk+e\bA$ with
$A=(0, Bx, 0)$, the Hamiltonian in Eq.~(\ref{dw}) will be rewritten as
\begin{eqnarray}
H_{\chi}(\mathbf{\Pi})=\lambda(\Pi_{x}^{2}-\Pi_{y}^{2})\tau_{x}+\lambda(\Pi_{x}\Pi_{y}+\Pi_{y}\Pi_{x})\tau_{y}
+\chi v\Pi_{z}.\label{mdw}
\end{eqnarray}
Note that $\Pi_{x}$ and $\Pi_{y}$ do not commute, so we have adopted the symmetric ordering
for operators. The magnetic field will discretize the continuum energy spectra into
a series of Landau levels (or Landau bands), whose form can be easily obtained
by introducing the following ladder operators,
\begin{eqnarray}
a = \frac{l_B}{\sqrt{2}}(\Pi_x-i\Pi_y),\qquad a^{\dagger} = \frac{l_B}{\sqrt{2}}(\Pi_x+i\Pi_y),\label{ladder}
\end{eqnarray}
where $l_{B}=1/\sqrt{eB}$ is the magnetic length. Bring Eq.~(\ref{ladder})
into Eq.~(\ref{mdw}) and rewrite the Hamiltonian into a matrix form,
\begin{eqnarray}
    H = \left(\begin{array}{cc}
 \chi vk_{z} & \frac{2\lambda}{l_{B}^{2}}  a^2\\
 \frac{2\lambda}{l_{B}^{2}} {a^{\dagger}}^2 & -\chi vk_{z} \\
\end{array}
\right),
\end{eqnarray}
it is readily found that the Landau levels are given by
\begin{eqnarray}
E_{n,s,\chi}(k_{z})=\left\{\begin{array}{cc}
          s E_{n}, & n\geq2, \\
          -\chi vk_{z}, & n=0,1,
        \end{array}\right.
\end{eqnarray}
where $E_n(k_z) \equiv \sqrt{v^{2}k_{z}^{2}+4n(n-1)(\lambda eB)^{2}}$; $s=+$ and $s=-$ correspond to the particle branch and hole branch, respectively; and
when $n=0$ or $1$, $s=+$ is implicitly assumed.
A graph illustration of the Landau levels is shown in Fig.~\ref{LL}.
Before proceeding, we make two remarks on the Landau levels:
(i) $E_{n}(0)\propto B$,  such a linear dependence is distinct from the $E_{n}(0)\propto \sqrt{B}$ law found in the Dirac and single Weyl semimetals~\cite{Gusynin2007MOC,Ashby2013weyl};
(ii) there are two degenerate chiral Landau levels ($n=0, 1$), with the number equating to the monopole charge of the Weyl points.
As we will show in Sec.~\ref{sub:phsb}, when the degeneracy is lifted, optical transition between the two chiral Landau levels will take place. 

For $n\geq2$, we denote the eigenvectors corresponding
to $E_{n,s,\chi}(k_{z})$ as $|n,s,\chi\rangle=(\alpha_{n,s,\chi}|n-2\rangle,\beta_{n,s,\chi}|n\rangle)^{T}$,
where $|n\rangle$ satisfies $a|n\rangle=\sqrt{n}|n-1\rangle$ and $a^{\dag}|n\rangle=\sqrt{n+1}|n+1\rangle$.
The coefficients $\alpha_{n,s,\chi}$ and $\beta_{n,s,\chi}$ can be determined straightforwardly, with
\begin{eqnarray}
\alpha_{n,s,\chi}=s \sqrt{\frac{E_{n,s,\chi}+\chi vk_{z}}{2E_{n,s,\chi}}},\quad
\beta_{n,s,\chi}=\sqrt{\frac{E_{n,s,\chi}-\chi vk_{z}}{2E_{n,s,\chi}}}.
\end{eqnarray}
For $n=0,1$, the eigenvectors are simply $(0,1)^{T}$.


The optical conductivity can be obtained from the Kubo formula, which is
\begin{eqnarray}
\sigma_{\mu \nu} (\omega) &=& \frac{-i}{2 \pi l_B^2} \sum_{n,n',s,s',\chi}
\int \frac{dk_z}{2\pi} \frac{f_{n,s,\chi}-f_{n',s',\chi}}{E_{n,s,\chi}-E_{n',s',\chi}} \nonumber\\
&&\times\frac{\langle n,s,\chi|j_{\mu,\chi}|n',s',\chi\rangle\langle n',s',\chi|j_{\nu,\chi}|n,s,\chi\rangle}{\omega-E_{n,s,\chi}+E_{n',s',\chi}+i\Gamma},\label{kubo}
\end{eqnarray}
where $f_{n,s,\chi}=1/[\exp{(E_{n,s,\chi}-\mu)/T}+1]$ is the Fermi-Dirac distribution function,
with $\mu$ the chemical potential and $T$ the temperature; $\Gamma$ denotes the impurity scattering rate,
in this work, we assume that all Landau levels share the same $\Gamma$ for simplicity;
$j_{\mu,\chi}$ ($\mu=x,y,z$) denote the current operators, whose explicit forms are
\begin{eqnarray}
    j_{x,\chi} &=& \frac{\partial H_{\chi}}{\partial A_x} = 2 e \lambda \left[ \Pi_x \sigma_x+\Pi_y \sigma_y \right] \nonumber = 2\sqrt{2}\frac{e\lambda}{l_B} \left(
\begin{array}{cc}
0 & a\\
a^{\dagger} & 0 \\
\end{array}
\right), \\
    j_{y,\chi} &=& \frac{\partial H_{\chi}}{\partial A_y} = 2 e \lambda \left[-\Pi_y \sigma_x+\Pi_x \sigma_y \right] \nonumber = i2\sqrt{2}\frac{e\lambda}{l_B}  \left(
\begin{array}{cc}
0 & -a\\
a^{\dagger} & 0 \\
\end{array}
\right), \\
    j_{z,\chi} &=& \frac{\partial  H_{\chi}}{\partial A_z} = \chi e v  \sigma_z = \chi e v  \left(
\begin{array}{cc}
1 & 0\\
0 & -1\\
\end{array}
\right).\label{current}
\end{eqnarray}

In the clean limit, i.e., $\Gamma=0$,  the dissipative components
which correspond to the absorption of light are found to be
\begin{widetext}
\begin{eqnarray}
\Re(\sigma_{xx}(\omega))&=&-\frac{4e^{2}\lambda^{2}}{l_{B}^{4}} \sum_{n=1} n \int \frac{dk_z}{2 \pi} \left[\frac{f_{n,+}-f_{n,-}-f_{n+1,+}+f_{n+1,-}}{E_{n}-E_{n+1}} \left(1-\frac{v^{2}k_z^2}{E_{n} E_{n+1}}\right) \delta \left( \omega +E_{n}-E_{n+1} \right)\right. \nonumber\\
&&\left.+ \frac{f_{n,+}-f_{n,-}+f_{n+1,+}-f_{n+1,-}}{E_{n}+E_{n+1}}
\left(1+\frac{v^2 k_z^2}{E_{n} E_{n+1}}\right) \delta \left( \omega -E_{n}-E_{n+1} \right)
\right],\nonumber
\end{eqnarray}
\end{widetext}

\begin{widetext}
\begin{eqnarray}
\Re({\sigma_{zz}(\omega)})&=&-\frac{e^2 v ^{2}}{ l_{B}^2} \sum_{n=2} \int \frac{dk_{z} }{2 \pi} \frac{f_{n,+}-f_{n,-}}{2 E_{n}} (1-\frac{v^{2}k_{z}^2}{E_{n}^2}) \delta \left( \omega -2E_{n} \right),\nonumber\\
\Im(\sigma_{xy}(\omega))&=&-\frac{4e^{2}\lambda^{2}}{l_{B}^{4}}\sum_{n=1} n \int \frac{dk_z}{2 \pi}\left[ \frac{-f_{n,+}-f_{n,-}+f_{n+1,+}+f_{n+1,-}}{E_{n}-E_{n+1}} \left(1-\frac{v^{2}k_z^2}{E_{n} E_{n+1}}\right) \delta \left( \omega +E_{n}-E_{n+1} \right)\right. \nonumber\\
&&\left.+ \frac{f_{n,+}+f_{n,-}-f_{n+1,+}-f_{n+1,-}}{E_{n}+E_{n+1}} \left(1+\frac{v^{2}k_z^2}{E_{n} E_{n+1}}\right) \delta \left( \omega -E_{n}-E_{n+1} \right)\right],\label{transition}
\end{eqnarray}
\end{widetext}
where we have defined $f_{n,\pm} \equiv 1/[\exp{(\pm E_{n}-\mu)/T}+1]$ for simplicity of notation.

Note that $\Re(\sigma_{yy})=\Re(\sigma_{xx})$ due to rotational symmetry. From Eq.~(\ref{kubo}) and Eq.~(\ref{current}),
it is readily found that the other two transverse components $\sigma_{x z}$ and $\sigma_{y z}$ are zero.
The selection rules of the
optical transition can be immediately read out from the $\delta$-functions in Eq.~(\ref{transition}).
The $\delta$-functions in the first line of $\Re(\sigma_{xx})$ and $\Im(\sigma_{xy})$ denote the transitions between nearest-neighbor Landau levels within the same branch, i.e.,
the selection rules are $\Delta n=n_{p,f}-n_{p,i}=1$ and $\Delta n=n_{h,f}-n_{h,i}=-1$, where $n_{p(h),i}$ and $n_{p(h),f}$
denote the indices of the initial particle (hole) Landau level and the final
particle (hole) Landau level of the transition process, respectively (see the caption of Fig.~\ref{LL}).
The $\delta$-functions in the second line of $\Re(\sigma_{xx})$ and $\Im(\sigma_{xy})$ indicate
that the transitions can also occur between the particle-branch Landau levels and the hole-branch Landau levels,
with the selection rules $\Delta n=n_{p,f}-n_{h,i}=\pm1$.
The $\delta$-function in $\Re(\sigma_{zz})$ indicates that the transitions can only occur
between the particle-branch Landau levels and the hole-branch Landau levels, with the selection rule
$\Delta n=n_{p,f}-n_{h,i}=0$.

After performing the integration over $k_{z}$, we find that
\begin{widetext}
\begin{eqnarray}
\Re(\sigma_{xx}(\bar\omega))&=&\frac{2 e^{2}\lambda }{\pi v l_{B}^{2}} \sum_{n=1}^{\xi(\bar \omega)}  \left[-\frac{\sinh \left(\frac{2 n-\bar\omega ^2}{2 \bar{T} \omega }\right)}{\cosh \left(\frac{2 n-\bar\omega ^2}{2 \bar{T} \bar\omega }\right)+\cosh \left(\frac{\bar\mu }{\bar{T}}\right)} + \frac{\sinh \left(\frac{2 n+\bar\omega ^2}{2 \bar{T} \omega }\right)}{\cosh \left(\frac{2 n+\bar\omega ^2}{2 \bar{T} \bar\omega }\right)+\cosh \left(\frac{\bar\mu }{\bar{T}}\right)} \right] \frac{n}{\bar\omega} \frac{|2 n^2 - \bar\omega ^2|}{\sqrt{4n^2-4n^2\bar \omega^2+\bar \omega^4}}  \theta(|\sqrt{2n}-\bar\omega|),\nonumber\\
\Re(\sigma_{zz}(\bar\omega))&=&\frac{e^2 v }{2\pi \lambda}  \frac{\sinh \left(\frac{\bar\omega }{2 \bar{T}}\right)}{\cosh \left(\frac{\bar \mu }{\bar{T}}\right)+\cosh \left(\frac{\bar\omega }{2 \bar{T}}\right)} \sum_{n=2}^{\lfloor\frac{1+\sqrt{\bar\omega ^2+1}}{2}\rfloor}  \frac{n^2-n}{\bar\omega^2\sqrt{4n-4n^2+\bar \omega^2}},\nonumber\\
\Im(\sigma_{xy}(\bar\omega))&=&-\frac{2e^{2}\lambda}{\pi v l_{B}^{2}} \sum_{n=1}^{\xi(\bar \omega)}  \left[\frac{\cosh \left(\frac{2 n-\bar\omega ^2}{2 \bar{T} \omega }\right)+\exp\left({\frac{\bar\mu }{\bar{T}}}\right)}{\cosh \left(\frac{2 n-\bar\omega ^2}{2 \bar{T} \bar\omega }\right)+\cosh \left(\frac{\bar\mu }{\bar{T}}\right)}  -  \frac{\cosh \left(\frac{2 n+\bar\omega ^2}{2 \bar{T} \omega }\right)+\exp\left({\frac{\bar\mu }{\bar{T}}}\right)}{\cosh \left(\frac{2 n+\bar\omega ^2}{2 \bar{T} \bar\omega }\right)+\cosh \left(\frac{\bar \mu }{\bar{T}}\right)} \right] \frac{n}{\bar\omega} \frac{|2 n^2 - \bar\omega ^2|}{\sqrt{4n^2-4n^2\bar \omega^2+\bar \omega^4}}  \theta(|\sqrt{2n}-\bar\omega|),\label{expression}
\end{eqnarray}
\end{widetext}
where $\bar{\omega}\equiv \omega l_{B}^{2}/2\lambda$, $\bar{\mu}\equiv \mu l_{B}^{2}/2\lambda$
and $\bar{T}\equiv T l_{B}^{2}/2\lambda$ are dimensionless quantities, and the upper bound of the
summation over $n$, $\xi(\bar \omega)$, is given by
\begin{eqnarray}
    \xi(\bar{\omega})=
\left\{\begin{array}{cc}
                      \lfloor\frac{\bar{\omega} ^2}{2 \sqrt{\bar{\omega} ^2-1}}\rfloor, & \text{if } \bar{\omega} > 1, \\
                      \infty, & \text{otherwise.}
                    \end{array}\right.
\end{eqnarray}
The expressions of the optical conductivity are too complicated to obtain
analytical results. Thus, we will resort to numerical calculations in the following discussions.

\section{Magneto-optical conductivity}
\label{MOC}
\subsection{Longitudinal component}

\begin{figure}
\subfigure{\includegraphics[width=7cm, height=6cm]{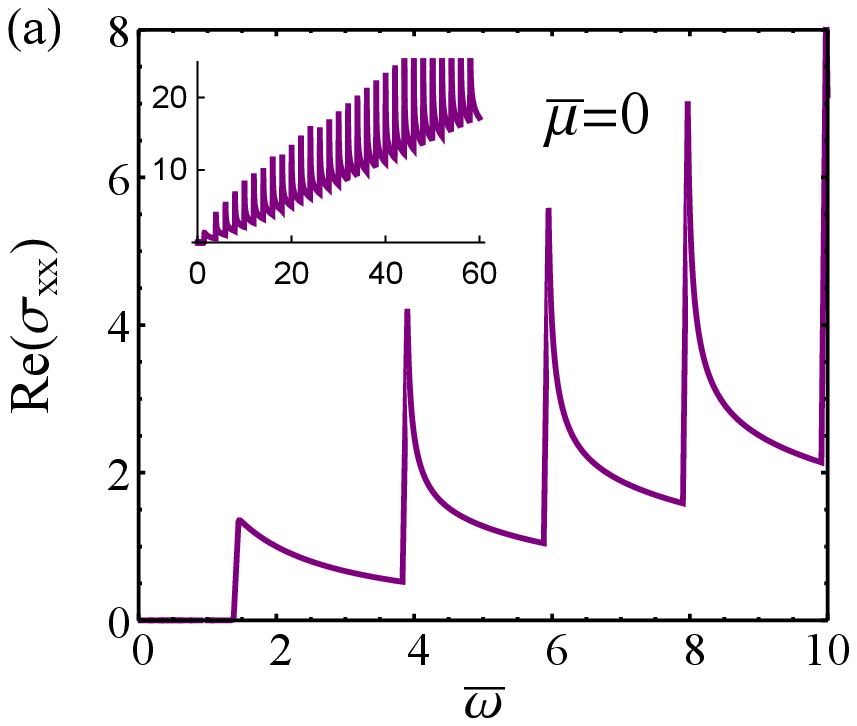}}
\subfigure{\includegraphics[width=7cm, height=6cm]{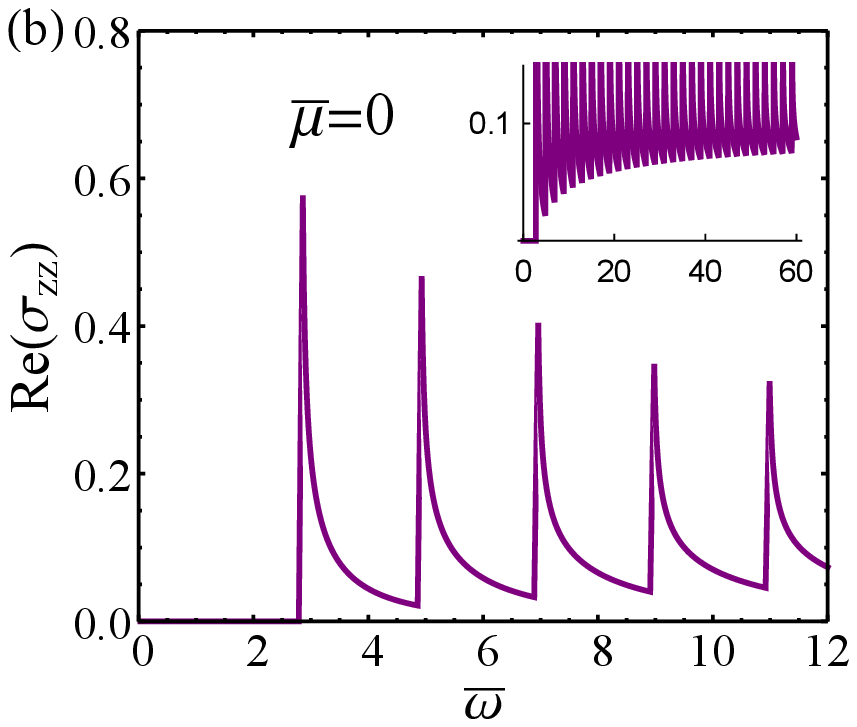}}
\caption{The real part of the longitudinal conductivity at $\mu=0$. Here $T l_{B}^{2}/2\lambda = 0.01$, $\Gamma=0$.
(a) For $\text{Re}(\sigma_{xx})$ (in units of $2e^2\lambda / \pi v  l_B^2$), a series of resonant peaks are observed at $\bar{\omega}=\sqrt{n(n+1)}+\sqrt{n(n-1)}$.
(b) For $\text{Re}(\sigma_{zz})$ (in units of $e^2 v  / 2 \pi \lambda$), the peaks are located at $\bar{\omega}=2\sqrt{n(n-1)}$.
A broader frequency range of the optical conductivity is shown in the insets, from which
it is readily seen that for $\text{Re}(\sigma_{xx})$,
the minima of the optical conductivity exhibit a linear dependence on
the frequency, while for $\text{Re}(\sigma_{zz})$, the minima of the optical conductivity approach a constant for larger frequency.} \label{LMOC}
\end{figure}

Fig.~\ref{LMOC} shows the real part of the longitudinal  conductivity at the neutrality
condition, i.e., $\mu=0$. For both $\text{Re}(\sigma_{xx})$ and $\text{Re}(\sigma_{zz})$,
the most prominent feature is a series of resonant peaks originated from the optical transitions
between Landau levels.
Due to the difference in selection rules, it is found that
the resonant peaks for $\text{Re}(\sigma_{xx})$ are located at
$\omega=2[\sqrt{n(n+1)}+\sqrt{n(n-1)}]\lambda eB$, while for $\text{Re}(\sigma_{zz})$,
they are located at $\omega=4\sqrt{n(n-1)}\lambda eB$. It is immediately seen
that the frequencies of the resonant peaks, for both $\text{Re}(\sigma_{xx})$ and $\text{Re}(\sigma_{zz})$,
follow an $\omega\varpropto B$ law, which is distinct from the
$\omega\varpropto \sqrt{B}$ law obeyed by the Dirac semimetals and single Weyl
semimetals~\cite{Gusynin2007MOC, Ashby2013weyl}.
In experiments, the frequency ratio between neighboring peaks is an important quantity
for figuring out the Landau level indices related to the peaks, and further extracting information about the underlying band structure, e.g., whether an energy gap
exists or not, and the value of Fermi velocity~\cite{Jiang2007moc,Deacon2007moc, Orlita2015MOC,Chen2016ZrTe5,Akrap2016MOC}.
Here the frequency ratio between the $(n+1)$-th and $n$-th (counting from the left) resonant peaks  for
$\text{Re}(\sigma_{xx})$ is given by  $[\sqrt{(n+2)(n+1)}+\sqrt{n(n+1)}]/[\sqrt{n(n+1)}+\sqrt{n(n-1)}]$,
and for $\text{Re}(\sigma_{zz})$ it is $\sqrt{n+2}/\sqrt{n}$. Furthermore,
the energy offset between the $(n+1)$-th and $n$-th peaks is found to decrease with
the increase of $n$ and saturates to $4\lambda eB$ for large $n$,
such a behavior is also distinct from the Dirac and single Weyl semimetals, in which the energy offset will
monotonically decrease to zero~\cite{Gusynin2007MOC, Ashby2013weyl}.

Besides the difference in the locations of the resonant peaks, another distinctive feature
between $\text{Re}(\sigma_{xx})$ and $\text{Re}(\sigma_{zz})$ is that
the resonant peaks of the former sit on a linear background,  while those of the latter sit on
a constant background (see insets of Fig.~\ref{LMOC}). Both observations agree with the fact that without a magnetic field, the optical conductivity
$\text{Re}(\sigma_{xx})$ is linear in frequency whereas $\text{Re}(\sigma_{zz})$ is a constant independent of frequency~\cite{Ahn2017mw}. 
Such distinction originates from the apparent discrepancy of dispersion relations between $x$-$y$ plane and $z$ direction and is not present in single Weyl semimetals for which all longitudinal components should
reside at a linear background~\cite{Ashby2013weyl}.

\begin{figure}[t]
\subfigure{\includegraphics[width=7cm, height=6cm]{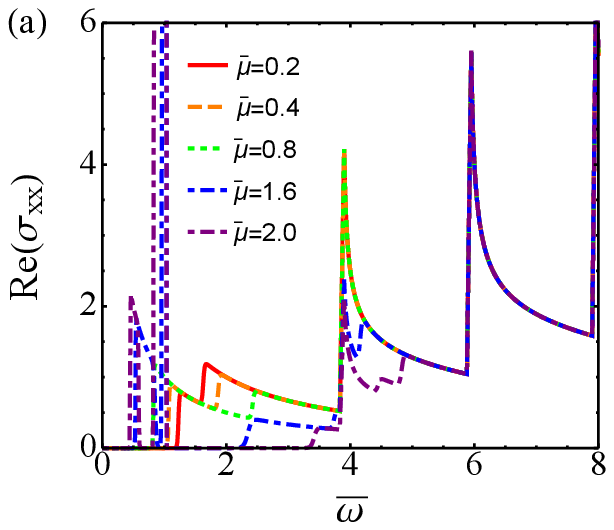}}
\subfigure{\includegraphics[width=7cm, height=6cm]{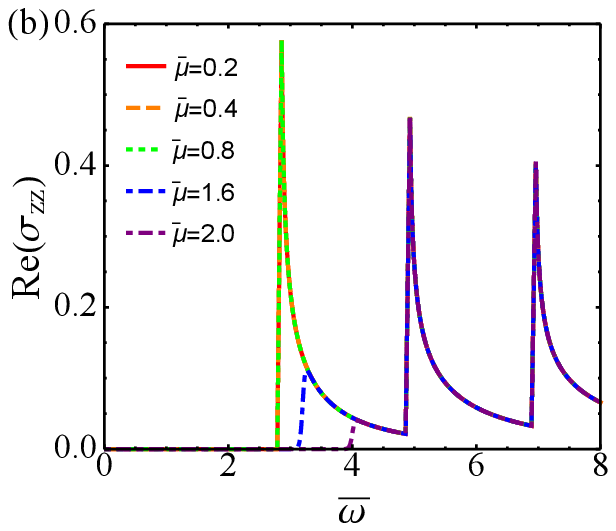}}
\caption{The effect of chemical potential to the real part of longitudinal conductivity. Here $T l_{B}^{2}/2\lambda = 0.01$, $\Gamma=0$.
(a) For $\text{Re}(\sigma_{xx})$ (in units of $2e^2\lambda/\pi v  l_B^2$),  the absorption edges related to the chiral Landau levels are
moving with the variation of $\mu$. When $\mu$ crosses the minimum of a Landau level, a new resonant peak shows up. (b) For
 $\text{Re}(\sigma_{zz})$ (in units of $e^2 v  / 2 \pi \lambda$),  when $\mu$ crosses the minimum of a Landau level, the resonant peak related to the
 Landau level changes to an absorption edge whose position is directly determined by $\mu$.
} \label{ce}
\end{figure}

Fig.~\ref{ce} further illustrates the effect of chemical potential $\mu$ to the longitudinal
conductivity. For $\text{Re}(\sigma_{xx})$, when $\mu$ is tuned away from the neutrality
condition but still only crosses the chiral Landau levels, i.e., $\bar{\mu}<\sqrt{2}$ (due to particle-hole symmetry, we will only consider the $\mu>0$ case),
the relatively weak peak at $\bar{\omega}=\sqrt{2}$
shown in Fig.~\ref{LMOC}(a) will be split into two absorption edges which are located at
$\bar{\omega}_{1}=\sqrt{\bar{\mu}^{2}+2}-\bar{\mu}$ and $\bar{\omega}_{2}=\sqrt{\bar{\mu}^{2}+2}+\bar{\mu}$,
as shown in Fig.~\ref{ce}(a). For $\text{Re}(\sigma_{zz})$, when $\mu$ only crosses the chiral Landau levels, however,
the change of $\mu$ has no effect to the optical conductivity, this is because the selection rule for
$\text{Re}(\sigma_{zz})$ is $\Delta n=n_{p,f}-n_{h,i}=0$ with $n_{p(h)}\geq2$. When $\mu$ is tuned to cross the $n$-th
particle Landau level with $n\geq2$,
we find that for $\text{Re}(\sigma_{xx})$, a new resonant peak will appear
at $\bar{\omega}=\sqrt{n(n+1)}-\sqrt{n(n-1)}$ (the case with $n=2$ is shown in Fig.~\ref{ce}(a)).
The new peak corresponds to the intra-branch transition between the
$n$-th and $(n+1)$-th particle Landau levels. The transition is now allowed because the $n$-th particle Landau  level gets occupied
while the  $(n+1)$-th particle Landau level remains empty. For $\text{Re}(\sigma_{zz})$, when $\mu$ is tuned to cross the $n$-th particle Landau level,
the resonant peak at $\bar{\omega}=2\sqrt{n(n-1)}$ will disappear and an absorption edge will appear at
$\bar{\omega}=2\bar{\mu}$ (the case with $n=2$ is shown in Fig.~\ref{ce}(b)).

\subsection{Transverse component}

\begin{figure}[t]
\subfigure{\includegraphics[width=7cm, height=6cm]{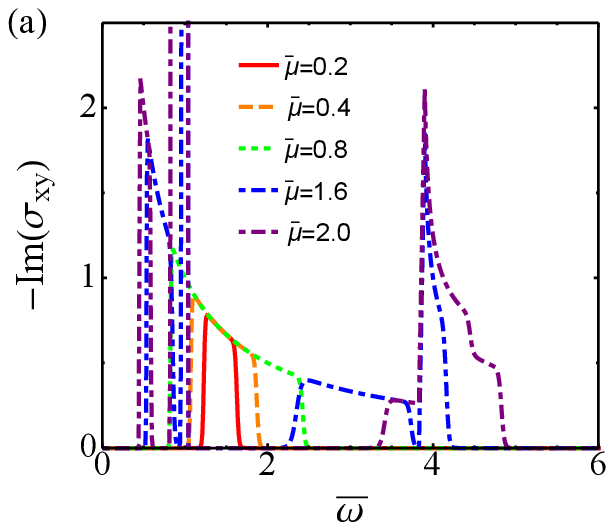}}
\subfigure{\includegraphics[width=7cm, height=6cm]{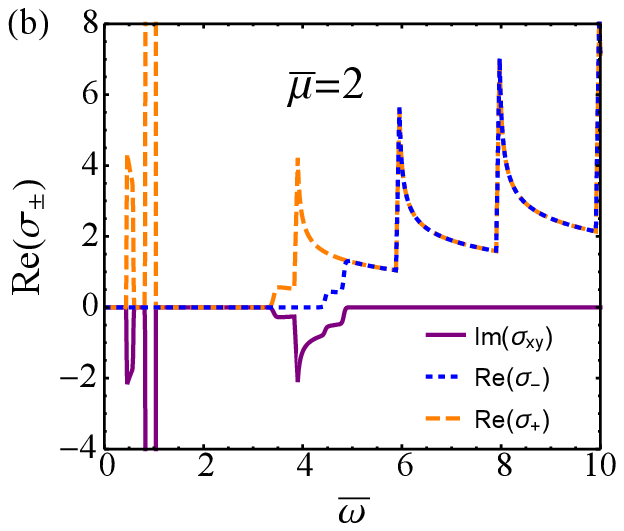}}
\caption{The absorption part of the transverse conductivity  (in units of $2e^2\lambda/\pi v  l_B^2$).
Here $T l_{B}^{2}/2\lambda = 0.01$, $\Gamma=0$.
(a) $\text{Im}(\sigma_{xy})$ for
several values of $\mu$. (b) The absorption part of the optical
conductivity for circularly polarized light. } \label{TMOC}
\end{figure}

The transverse (Hall) conductivity $\text{Im}(\sigma_{xy})$ at $\mu=0$ vanishes for
all frequencies, this fact can be read directly  from the expression of $\text{Im}(\sigma_{xy})$
in Eq.~(\ref{expression}). A more physical explanation is that the transverse conductivity has
four contributions: I, the transition from the $n$-th hole-branch Landau level to
the $(n+1)$-th particle-branch Landau level; II, the transition from the
$(n+1)$-th hole-branch Landau level to the $n$-th particle-branch Landau level; III, the transition
from the $(n+1)$-th hole-branch Landau level to the $n$-th hole-branch Landau level;
IV, the transition
from the $n$-th particle-branch Landau level to the $(n+1)$-th particle-branch Landau level.
Importantly, the contributions from I and III have a sign difference to the ones from II and IV,
respectively. Moreover, at $\mu=0$, due to particle-hole symmetry the contributions exactly cancel each other out;
thus, $\text{Im}(\sigma_{xy})$ vanishes for all frequencies in this case.
Away from the neutrality condition, particle-hole symmetry is broken and consequently the transverse
conductivity $\text{Im}(\sigma_{xy})$
will take nonzero values in certain frequency regions, as shown in Fig.~\ref{TMOC}(a).

In experiments, the quantities $\sigma_{\pm}\equiv\sigma_{xx}\pm i\sigma_{xy}$ are also
of interest because they can be used to determine the polarization of lights. Concretely,
$\sigma_{+}$ corresponds to right-handed polarized light, and $\sigma_{-}$
corresponds to left-handed polarized light. The absorption parts of $\sigma_{\pm}$
are shown in Fig.~\ref{TMOC}(b). At the lower frequency regime, i.e., $\omega\lesssim2\mu$,
$\text{Re}(\sigma_{-})$ vanishes as the longitudinal conductivity makes a cancellation
with the transverse conductivity. The picture is similar to that of the single Weyl semimetals~\cite{Ashby2013weyl}.

\subsection{Temperature effect and impurity scattering effect}

\begin{figure}[t]
\subfigure{\includegraphics[width=7cm, height=6cm]{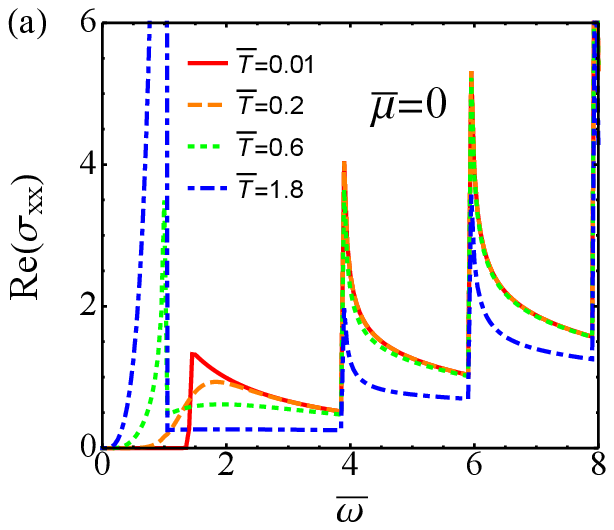}}
\subfigure{\includegraphics[width=7cm, height=6cm]{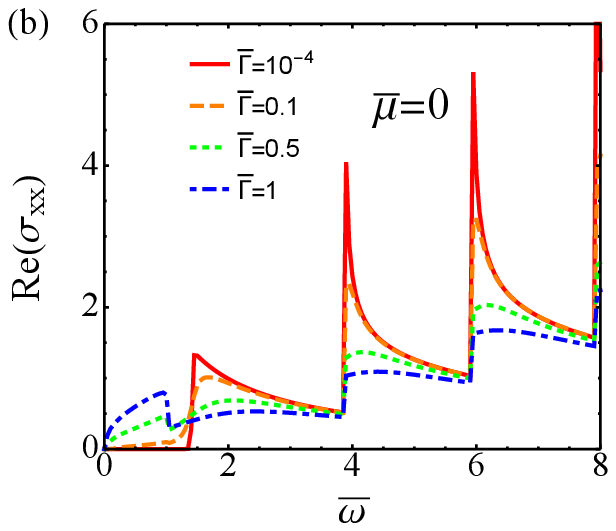}}
\caption{The effect of temperature and impurity scattering to the
optical conductivity (in units of $2e^2\lambda/\pi v  l_B^2$). (a) is plotted in $\Gamma l_{B}^{2}/2\lambda=10^{-4}$, (b) in $T l_{B}^{2}/2\lambda = 0.01$. A suppression of resonant peaks is observed in both cases.} \label{tem}
\end{figure}

The effect of finite temperature to the optical conductivity enters through the Fermi-Dirac
distribution function. Fig.~\ref{tem}(a) shows the temperature effect to $\text{Re}(\sigma_{xx})$ at
$\mu=0$,
it is readily seen that with the increase of temperature, the height of the resonant peaks
will be suppressed, and resonant peaks corresponding to intra-branch transition which
are absent at zero temperature will show up
(see the blue dashed line and the green dashed line). The effect of impurity scattering to
$\text{Re}(\sigma_{xx})$ is shown in Fig.~\ref{tem}(b). It is readily seen that
the main effect of the impurity scattering is just a suppression
of the resonant peaks.

For $\text{Re}(\sigma_{zz})$
under the neutrality condition, finite temperature and impurity scattering
mainly induce a suppression to the resonant peaks, no new peaks will show up
(not shown here).

\subsection{Effect of particle-hole symmetry breaking} 
\label{sub:phsb}



Next we will explore the effect of particle-hole symmetry breaking since in real materials there is no constraint to preserve this symmetry. To achieve this, we add a diagonal quadratic term of the form $\frac{k_{x}^2+k_{y}^{2}}{2m} \tau_0$ to the original Hamiltonian in Eq.~(\ref{dw}), with $m \gtrsim \lambda^{-1}$. The Hamiltonian with standard minimum coupling in matrix form thus changes to

\begin{eqnarray}
    H = \left(\begin{array}{cc}
 \chi vk_{z} + sb & \frac{2\lambda}{l_{B}^{2}}  a^2\\
 \frac{2\lambda}{l_{B}^{2}} {a^{\dagger}}^2 & -\chi vk_{z}+sb \\
\end{array}
\right),
\end{eqnarray}
with the symmetry-breaking term $sb= \omega_{c}(a^{\dag}a+\frac{1}{2})$, where $\omega_{c}=eB/m$.
Correspondingly, the Landau levels become
\begin{eqnarray}
E_{n,s,\chi}(k_{z})&=&s\sqrt{(\chi vk_{z}-\omega_{c})^{2}+4n(n-1)(\lambda eB)^{2}}\nonumber\\
&&+(n-\frac{1}{2})\omega_{c},\quad n\geq2,\nonumber\\
E_{1,s,\chi}(k_{z})&=&-\chi vk_{z}+\frac{3}{2}\omega_{c},\nonumber\\
E_{0,s,\chi}(k_{z})&=&-\chi vk_{z}+\frac{1}{2}\omega_{c}.
\end{eqnarray}
As the symmetry-breaking term  does not change the selection rules
(this can be simply confirmed by bringing the current operators into the Kubo formula),
the positions of the resonant peaks can be readily figured out. For $\text{Re}(\sigma_{zz})$,
the positions of the resonant peaks do no change. While for $\text{Re}(\sigma_{xx})$, the main resonant
peaks will undergo splitting and position shift. Concretely, the inter-branch $\omega=2[\sqrt{n(n+1)}+\sqrt{n(n-1)}]\lambda eB$ peaks will be split to $\omega= \pm \omega_{c} + 2[\sqrt{(n+1)n}+\sqrt{n(n-1)}]\lambda eB$, and the intra-branch $\omega=2[\sqrt{n(n+1)}-\sqrt{n(n-1)}]\lambda eB$ peaks will be split to $\omega= \pm\omega_{c} + 2[\sqrt{(n+1)n}-\sqrt{n(n-1)}]\lambda eB$.


More interestingly, as the degeneracy of the two chiral Landau levels is
lifted by the symmetry-breaking term, an optical transition within the two chiral Landau levels is allowed.
As the energy offset between the two chiral Landau levels is fixed to $\omega_{c}$ for arbitrary $k_{z}$,
an absorption peak at $\omega=\omega_{c}$ is expected.  Compared to the transitions between 
chiral Landau levels and non-chiral Landau levels, this transition is remarkable in the sense
that it will not be influenced by the change of the chemical potential as long as 
the chiral Landau levels cross the Fermi level (if we restrict ourselves to the continuum model, this 
is always the case).
Note that this effect is exclusive to multi-Weyl semimetals as there is no chiral Landau level degeneracy to be lifted for singe-Weyl semimetals.


Although we have chosen a simple diagonal term to discuss the effect of
particle-hole symmetry breaking, the above conclusions will qualitatively
hold for more complicated symmetry-breaking terms.
From the results above, we can see that although the particle-hole symmetry breaking will affect
the frequencies of the resonant peaks, the influence is in fact
marginal and can be analyzed and regulated in real experiments.
The linear scaling dependence
on the magnetic field strength and the anisotropic behavior
of the longitudinal conductivity still hold.

\subsection{Generalization to higher monopole charge}

The above discussions could be systematically generated to multi-Weyl semimetals with arbitrary monopole charge $Q$. Specifically,
the number of  chiral Landau levels equals $Q$, and the Landau levels behave as $E_n(0)\propto B^{Q/2}$; while the selection rules for
optical conductivity remains unmodified.

Thus for $Q \geq 3$, the scaling dependence of the resonant-peak frequencies on the magnetic field will follow
the law $\omega\propto B^{Q/2}$, and the longitudinal conductivity will exhibit more anisotropic
behaviors. This general observation can be utilized to determine the monopole charge of multi-Weyl semimetals.

\section{Experimental estimation}
\label{ee}

Now we give a brief experimental estimation. We take the material candidate HgCr$_{2}$Se$_{4}$~\cite{Xu2011doubleweyl,Fang2012MW}
as a concrete example. It was predicted that there are only two double-Weyl points in HgCr$_{2}$Se$_{4}$ which are very close
to the Fermi energy~\cite{Xu2011doubleweyl,Fang2012MW}, thus, it is an ideal platform to test the above
predictions on the optical conductivity of double-Weyl semimetals.

According to the band structure
calculated in Ref.~\cite{Fang2012MW},
we find $\lambda(\hbar\pi/a)^{2}\approx0.4$ eV with the lattice constant $a = 10.753$ {\AA}~\cite{baltzer1965}, which gives
$\lambda^{-1} \approx 1.63 m_e$. Then
the interval between two near-neighbor resonant peaks is estimated to
be $\Delta\omega\approx4\lambda \hbar eB \approx 0.28$ meV
for $B=1$ T. To make the resonant peaks well separated in energy, the magnetic field
can be chosen to the order of $10$ T. In accordance with the magnetic field,
the adequate light for the magneto-optical experiment falls into the mid-infrared regime.
As the energy scale is of the order of several meV,
the sample should better be very clean and the
experimental temperature should better be lower than $1$ meV, i.e., $T<10$ K.

\section{Discussions and conclusions}
\label{dc}

To conclude, in this paper, we studied the magneto-optical response of double Weyl semimetals.
We found that the longitudinal components of the optical conductivity have a series of resonant peaks
with their corresponding frequencies
scaling linearly with the strength of the magnetic field,  and their frequency ratios following
a special law determined by the underlying Landau level structure.
Note that the linear scaling law is distinct from the characteristic square-root
scaling law obeyed by the Dirac semimetals and single Weyl semimetals~\cite{Gusynin2007MOC, Ashby2013weyl}. We also
found that the optical conductivity is quite anisotropic, with two of the three longitudinal components
residing at a linear background and the rest one at a constant
background. The anisotropy of the optical conductivity simply originates from the anisotropy of
the energy dispersion of double Weyl semimetals. Because transitions related to the chiral Landau levels are strongly 
dependent on the positions of the Fermi energy, it is found that the tuning of chemical potential duly modifies the corresponding resonate peaks of the optical conductivity.

Furthermore, we found that 
when the degeneracy of the chiral Landau levels is lifted by a particle-hole symmetry
breaking term, a new transition channel within the chiral Landau levels is opened for $\text{Re}(\sigma_{xx})$ , with
the transition frequency directly controlled by symmetry breaking parameter and irresponsive to the change of chemical potential. 
The effect of
finite temperature and impurity scattering to the optical conductivity
is mainly a suppression of the resonant peaks, which indicates that our study is quite robust against these subtleties.

Among above results, the linear scaling law between the frequencies of the resonant peaks and the magnetic field,
and the anisotropic behavior of the longitudinal conductivity are two most distinctive features which can be
applied to determine whether the predicted candidate materials, like HgCr$_{2}$Se$_{4}$~\cite{Xu2011doubleweyl,Fang2012MW}
and SrSi$_{2}$~\cite{Huang2016double},
are truly double Weyl semimetals or not.

\section{Acknowledgments}

We are grateful to Zhongbo Yan and Phillip E. C. Ashby for their helpful discussions.
This work is supported by NSFC under Grant NO. 11375168.

\bibliography{dirac}

\begin{thebibliography}{64}%
\makeatletter
\providecommand \@ifxundefined [1]{%
 \@ifx{#1\undefined}
}%
\providecommand \@ifnum [1]{%
 \ifnum #1\expandafter \@firstoftwo
 \else \expandafter \@secondoftwo
 \fi
}%
\providecommand \@ifx [1]{%
 \ifx #1\expandafter \@firstoftwo
 \else \expandafter \@secondoftwo
 \fi
}%
\providecommand \natexlab [1]{#1}%
\providecommand \enquote  [1]{``#1''}%
\providecommand \bibnamefont  [1]{#1}%
\providecommand \bibfnamefont [1]{#1}%
\providecommand \citenamefont [1]{#1}%
\providecommand \href@noop [0]{\@secondoftwo}%
\providecommand \href [0]{\begingroup \@sanitize@url \@href}%
\providecommand \@href[1]{\@@startlink{#1}\@@href}%
\providecommand \@@href[1]{\endgroup#1\@@endlink}%
\providecommand \@sanitize@url [0]{\catcode `\\12\catcode `\$12\catcode
  `\&12\catcode `\#12\catcode `\^12\catcode `\_12\catcode `\%12\relax}%
\providecommand \@@startlink[1]{}%
\providecommand \@@endlink[0]{}%
\providecommand \url  [0]{\begingroup\@sanitize@url \@url }%
\providecommand \@url [1]{\endgroup\@href {#1}{\urlprefix }}%
\providecommand \urlprefix  [0]{URL }%
\providecommand \Eprint [0]{\href }%
\providecommand \doibase [0]{http://dx.doi.org/}%
\providecommand \selectlanguage [0]{\@gobble}%
\providecommand \bibinfo  [0]{\@secondoftwo}%
\providecommand \bibfield  [0]{\@secondoftwo}%
\providecommand \translation [1]{[#1]}%
\providecommand \BibitemOpen [0]{}%
\providecommand \bibitemStop [0]{}%
\providecommand \bibitemNoStop [0]{.\EOS\space}%
\providecommand \EOS [0]{\spacefactor3000\relax}%
\providecommand \BibitemShut  [1]{\csname bibitem#1\endcsname}%
\let\auto@bib@innerbib\@empty
\bibitem [{\citenamefont {Chiu}\ \emph {et~al.}(2016)\citenamefont {Chiu},
  \citenamefont {Teo}, \citenamefont {Schnyder},\ and\ \citenamefont
  {Ryu}}]{Chiu2015RMP}%
  \BibitemOpen
  \bibfield  {author} {\bibinfo {author} {\bibfnamefont {Ching-Kai}\
  \bibnamefont {Chiu}}, \bibinfo {author} {\bibfnamefont {Jeffrey C.~Y.}\
  \bibnamefont {Teo}}, \bibinfo {author} {\bibfnamefont {Andreas~P.}\
  \bibnamefont {Schnyder}}, \ and\ \bibinfo {author} {\bibfnamefont {Shinsei}\
  \bibnamefont {Ryu}},\ }\bibfield  {title} {\enquote {\bibinfo {title}
  {Classification of topological quantum matter with symmetries},}\ }\href
  {\doibase 10.1103/RevModPhys.88.035005} {\bibfield  {journal} {\bibinfo
  {journal} {Rev. Mod. Phys.}\ }\textbf {\bibinfo {volume} {88}},\ \bibinfo
  {pages} {035005} (\bibinfo {year} {2016})}\BibitemShut {NoStop}%
\bibitem [{\citenamefont {Murakami}(2007)}]{Murakami2007weyl}%
  \BibitemOpen
  \bibfield  {author} {\bibinfo {author} {\bibfnamefont {Shuichi}\ \bibnamefont
  {Murakami}},\ }\bibfield  {title} {\enquote {\bibinfo {title} {Phase
  transition between the quantum spin hall and insulator phases in 3d:
  emergence of a topological gapless phase},}\ }\href
  {http://stacks.iop.org/1367-2630/9/i=9/a=356} {\bibfield  {journal} {\bibinfo
   {journal} {New Journal of Physics}\ }\textbf {\bibinfo {volume} {9}},\
  \bibinfo {pages} {356} (\bibinfo {year} {2007})}\BibitemShut {NoStop}%
\bibitem [{\citenamefont {Wan}\ \emph {et~al.}(2011)\citenamefont {Wan},
  \citenamefont {Turner}, \citenamefont {Vishwanath},\ and\ \citenamefont
  {Savrasov}}]{wan2011}%
  \BibitemOpen
  \bibfield  {author} {\bibinfo {author} {\bibfnamefont {Xiangang}\
  \bibnamefont {Wan}}, \bibinfo {author} {\bibfnamefont {Ari~M.}\ \bibnamefont
  {Turner}}, \bibinfo {author} {\bibfnamefont {Ashvin}\ \bibnamefont
  {Vishwanath}}, \ and\ \bibinfo {author} {\bibfnamefont {Sergey~Y.}\
  \bibnamefont {Savrasov}},\ }\bibfield  {title} {\enquote {\bibinfo {title}
  {Topological semimetal and fermi-arc surface states in the electronic
  structure of pyrochlore iridates},}\ }\href {\doibase
  10.1103/PhysRevB.83.205101} {\bibfield  {journal} {\bibinfo  {journal} {Phys.
  Rev. B}\ }\textbf {\bibinfo {volume} {83}},\ \bibinfo {pages} {205101}
  (\bibinfo {year} {2011})}\BibitemShut {NoStop}%
\bibitem [{\citenamefont {Burkov}\ and\ \citenamefont
  {Balents}(2011)}]{burkov2011}%
  \BibitemOpen
  \bibfield  {author} {\bibinfo {author} {\bibfnamefont {A.~A.}\ \bibnamefont
  {Burkov}}\ and\ \bibinfo {author} {\bibfnamefont {Leon}\ \bibnamefont
  {Balents}},\ }\bibfield  {title} {\enquote {\bibinfo {title} {Weyl semimetal
  in a topological insulator multilayer},}\ }\href {\doibase
  10.1103/PhysRevLett.107.127205} {\bibfield  {journal} {\bibinfo  {journal}
  {Phys. Rev. Lett.}\ }\textbf {\bibinfo {volume} {107}},\ \bibinfo {pages}
  {127205} (\bibinfo {year} {2011})}\BibitemShut {NoStop}%
\bibitem [{\citenamefont {Weng}\ \emph {et~al.}(2015)\citenamefont {Weng},
  \citenamefont {Fang}, \citenamefont {Fang}, \citenamefont {Bernevig},\ and\
  \citenamefont {Dai}}]{weng2015}%
  \BibitemOpen
  \bibfield  {author} {\bibinfo {author} {\bibfnamefont {Hongming}\
  \bibnamefont {Weng}}, \bibinfo {author} {\bibfnamefont {Chen}\ \bibnamefont
  {Fang}}, \bibinfo {author} {\bibfnamefont {Zhong}\ \bibnamefont {Fang}},
  \bibinfo {author} {\bibfnamefont {B.~A.}\ \bibnamefont {Bernevig}}, \ and\
  \bibinfo {author} {\bibfnamefont {Xi}~\bibnamefont {Dai}},\ }\bibfield
  {title} {\enquote {\bibinfo {title} {Weyl semimetal phase in
  noncentrosymmetric transition-metal monophosphides},}\ }\href {\doibase
  10.1103/PhysRevX.5.011029} {\bibfield  {journal} {\bibinfo  {journal} {Phys.
  Rev. X}\ }\textbf {\bibinfo {volume} {5}},\ \bibinfo {pages} {011029}
  (\bibinfo {year} {2015})}\BibitemShut {NoStop}%
\bibitem [{\citenamefont {{Huang}}\ \emph {et~al.}(2015)\citenamefont
  {{Huang}}, \citenamefont {{Xu}}, \citenamefont {{Belopolski}}, \citenamefont
  {{Lee}}, \citenamefont {{Chang}}, \citenamefont {{Wang}}, \citenamefont
  {{Alidoust}}, \citenamefont {{Bian}}, \citenamefont {{Neupane}},
  \citenamefont {{Bansil}}, \citenamefont {{Lin}},\ and\ \citenamefont {{Zahid
  Hasan}}}]{Huang2015TaAs}%
  \BibitemOpen
  \bibfield  {author} {\bibinfo {author} {\bibfnamefont {S.-M.}\ \bibnamefont
  {{Huang}}}, \bibinfo {author} {\bibfnamefont {S.-Y.}\ \bibnamefont {{Xu}}},
  \bibinfo {author} {\bibfnamefont {I.}~\bibnamefont {{Belopolski}}}, \bibinfo
  {author} {\bibfnamefont {C.-C.}\ \bibnamefont {{Lee}}}, \bibinfo {author}
  {\bibfnamefont {G.}~\bibnamefont {{Chang}}}, \bibinfo {author} {\bibfnamefont
  {B.}~\bibnamefont {{Wang}}}, \bibinfo {author} {\bibfnamefont
  {N.}~\bibnamefont {{Alidoust}}}, \bibinfo {author} {\bibfnamefont
  {G.}~\bibnamefont {{Bian}}}, \bibinfo {author} {\bibfnamefont
  {M.}~\bibnamefont {{Neupane}}}, \bibinfo {author} {\bibfnamefont
  {A.}~\bibnamefont {{Bansil}}}, \bibinfo {author} {\bibfnamefont
  {H.}~\bibnamefont {{Lin}}}, \ and\ \bibinfo {author} {\bibfnamefont
  {M.}~\bibnamefont {{Zahid Hasan}}},\ }\bibfield  {title} {\enquote {\bibinfo
  {title} {{An inversion breaking Weyl semimetal state in the TaAs material
  class}},}\ }\href@noop {} {\bibfield  {journal} {\bibinfo  {journal} {Nat.
  Commun.}\ }\textbf {\bibinfo {volume} {6}},\ \bibinfo {pages} {7373}
  (\bibinfo {year} {2015})}\BibitemShut {NoStop}%
\bibitem [{\citenamefont {Xu}\ \emph {et~al.}(2015{\natexlab{a}})\citenamefont
  {Xu}, \citenamefont {Belopolski}, \citenamefont {Alidoust}, \citenamefont
  {Neupane}, \citenamefont {Bian}, \citenamefont {Zhang}, \citenamefont
  {Sankar}, \citenamefont {Chang}, \citenamefont {Yuan}, \citenamefont {Lee}
  \emph {et~al.}}]{Xu2015weyl}%
  \BibitemOpen
  \bibfield  {author} {\bibinfo {author} {\bibfnamefont {Su-Yang}\ \bibnamefont
  {Xu}}, \bibinfo {author} {\bibfnamefont {Ilya}\ \bibnamefont {Belopolski}},
  \bibinfo {author} {\bibfnamefont {Nasser}\ \bibnamefont {Alidoust}}, \bibinfo
  {author} {\bibfnamefont {Madhab}\ \bibnamefont {Neupane}}, \bibinfo {author}
  {\bibfnamefont {Guang}\ \bibnamefont {Bian}}, \bibinfo {author}
  {\bibfnamefont {Chenglong}\ \bibnamefont {Zhang}}, \bibinfo {author}
  {\bibfnamefont {Raman}\ \bibnamefont {Sankar}}, \bibinfo {author}
  {\bibfnamefont {Guoqing}\ \bibnamefont {Chang}}, \bibinfo {author}
  {\bibfnamefont {Zhujun}\ \bibnamefont {Yuan}}, \bibinfo {author}
  {\bibfnamefont {Chi-Cheng}\ \bibnamefont {Lee}},  \emph {et~al.},\ }\bibfield
   {title} {\enquote {\bibinfo {title} {Discovery of a weyl fermion semimetal
  and topological fermi arcs},}\ }\href@noop {} {\bibfield  {journal} {\bibinfo
   {journal} {Science}\ }\textbf {\bibinfo {volume} {349}},\ \bibinfo {pages}
  {613--617} (\bibinfo {year} {2015}{\natexlab{a}})}\BibitemShut {NoStop}%
\bibitem [{\citenamefont {Lv}\ \emph {et~al.}(2015)\citenamefont {Lv},
  \citenamefont {Weng}, \citenamefont {Fu}, \citenamefont {Wang}, \citenamefont
  {Miao}, \citenamefont {Ma}, \citenamefont {Richard}, \citenamefont {Huang},
  \citenamefont {Zhao}, \citenamefont {Chen} \emph {et~al.}}]{lv2015}%
  \BibitemOpen
  \bibfield  {author} {\bibinfo {author} {\bibfnamefont {BQ}~\bibnamefont
  {Lv}}, \bibinfo {author} {\bibfnamefont {HM}~\bibnamefont {Weng}}, \bibinfo
  {author} {\bibfnamefont {BB}~\bibnamefont {Fu}}, \bibinfo {author}
  {\bibfnamefont {XP}~\bibnamefont {Wang}}, \bibinfo {author} {\bibfnamefont
  {H}~\bibnamefont {Miao}}, \bibinfo {author} {\bibfnamefont {J}~\bibnamefont
  {Ma}}, \bibinfo {author} {\bibfnamefont {P}~\bibnamefont {Richard}}, \bibinfo
  {author} {\bibfnamefont {XC}~\bibnamefont {Huang}}, \bibinfo {author}
  {\bibfnamefont {LX}~\bibnamefont {Zhao}}, \bibinfo {author} {\bibfnamefont
  {GF}~\bibnamefont {Chen}},  \emph {et~al.},\ }\bibfield  {title} {\enquote
  {\bibinfo {title} {Experimental discovery of weyl semimetal taas},}\
  }\href@noop {} {\bibfield  {journal} {\bibinfo  {journal} {Phys. Rev. X}\
  }\textbf {\bibinfo {volume} {5}},\ \bibinfo {pages} {031013} (\bibinfo {year}
  {2015})}\BibitemShut {NoStop}%
\bibitem [{\citenamefont {Lu}\ \emph {et~al.}(2015)\citenamefont {Lu},
  \citenamefont {Wang}, \citenamefont {Ye}, \citenamefont {Ran}, \citenamefont
  {Fu}, \citenamefont {Joannopoulos},\ and\ \citenamefont {{Solja{\v
  c}i{\'c}}}}]{lu2015}%
  \BibitemOpen
  \bibfield  {author} {\bibinfo {author} {\bibfnamefont {Ling}\ \bibnamefont
  {Lu}}, \bibinfo {author} {\bibfnamefont {Zhiyu}\ \bibnamefont {Wang}},
  \bibinfo {author} {\bibfnamefont {Dexin}\ \bibnamefont {Ye}}, \bibinfo
  {author} {\bibfnamefont {Lixin}\ \bibnamefont {Ran}}, \bibinfo {author}
  {\bibfnamefont {Liang}\ \bibnamefont {Fu}}, \bibinfo {author} {\bibfnamefont
  {John~D.}\ \bibnamefont {Joannopoulos}}, \ and\ \bibinfo {author}
  {\bibfnamefont {Marin}\ \bibnamefont {{Solja{\v c}i{\'c}}}},\ }\bibfield
  {title} {\enquote {\bibinfo {title} {Experimental observation of weyl
  points},}\ }\href {\doibase 10.1126/science.aaa9273} {\ \textbf {\bibinfo
  {volume} {349}},\ \bibinfo {pages} {622--624} (\bibinfo {year}
  {2015})}\BibitemShut {NoStop}%
\bibitem [{\citenamefont {Young}\ \emph {et~al.}(2012)\citenamefont {Young},
  \citenamefont {Zaheer}, \citenamefont {Teo}, \citenamefont {Kane},
  \citenamefont {Mele},\ and\ \citenamefont {Rappe}}]{young2012dirac}%
  \BibitemOpen
  \bibfield  {author} {\bibinfo {author} {\bibfnamefont {Steve~M}\ \bibnamefont
  {Young}}, \bibinfo {author} {\bibfnamefont {Saad}\ \bibnamefont {Zaheer}},
  \bibinfo {author} {\bibfnamefont {Jeffrey~CY}\ \bibnamefont {Teo}}, \bibinfo
  {author} {\bibfnamefont {Charles~L}\ \bibnamefont {Kane}}, \bibinfo {author}
  {\bibfnamefont {Eugene~J}\ \bibnamefont {Mele}}, \ and\ \bibinfo {author}
  {\bibfnamefont {Andrew~M}\ \bibnamefont {Rappe}},\ }\bibfield  {title}
  {\enquote {\bibinfo {title} {Dirac semimetal in three dimensions},}\
  }\href@noop {} {\bibfield  {journal} {\bibinfo  {journal} {Phys. Rev. Lett.}\
  }\textbf {\bibinfo {volume} {108}},\ \bibinfo {pages} {140405} (\bibinfo
  {year} {2012})}\BibitemShut {NoStop}%
\bibitem [{\citenamefont {Wang}\ \emph {et~al.}(2012)\citenamefont {Wang},
  \citenamefont {Sun}, \citenamefont {Chen}, \citenamefont {Franchini},
  \citenamefont {Xu}, \citenamefont {Weng}, \citenamefont {Dai},\ and\
  \citenamefont {Fang}}]{wang2012dirac}%
  \BibitemOpen
  \bibfield  {author} {\bibinfo {author} {\bibfnamefont {Zhijun}\ \bibnamefont
  {Wang}}, \bibinfo {author} {\bibfnamefont {Yan}\ \bibnamefont {Sun}},
  \bibinfo {author} {\bibfnamefont {Xing-Qiu}\ \bibnamefont {Chen}}, \bibinfo
  {author} {\bibfnamefont {Cesare}\ \bibnamefont {Franchini}}, \bibinfo
  {author} {\bibfnamefont {Gang}\ \bibnamefont {Xu}}, \bibinfo {author}
  {\bibfnamefont {Hongming}\ \bibnamefont {Weng}}, \bibinfo {author}
  {\bibfnamefont {Xi}~\bibnamefont {Dai}}, \ and\ \bibinfo {author}
  {\bibfnamefont {Zhong}\ \bibnamefont {Fang}},\ }\bibfield  {title} {\enquote
  {\bibinfo {title} {Dirac semimetal and topological phase transitions in a 3
  bi (a= na, k, rb)},}\ }\href@noop {} {\bibfield  {journal} {\bibinfo
  {journal} {Phys. Rev. B}\ }\textbf {\bibinfo {volume} {85}},\ \bibinfo
  {pages} {195320} (\bibinfo {year} {2012})}\BibitemShut {NoStop}%
\bibitem [{\citenamefont {Wang}\ \emph {et~al.}(2013)\citenamefont {Wang},
  \citenamefont {Weng}, \citenamefont {Wu}, \citenamefont {Dai},\ and\
  \citenamefont {Fang}}]{wang2013three}%
  \BibitemOpen
  \bibfield  {author} {\bibinfo {author} {\bibfnamefont {Zhijun}\ \bibnamefont
  {Wang}}, \bibinfo {author} {\bibfnamefont {Hongming}\ \bibnamefont {Weng}},
  \bibinfo {author} {\bibfnamefont {Quansheng}\ \bibnamefont {Wu}}, \bibinfo
  {author} {\bibfnamefont {Xi}~\bibnamefont {Dai}}, \ and\ \bibinfo {author}
  {\bibfnamefont {Zhong}\ \bibnamefont {Fang}},\ }\bibfield  {title} {\enquote
  {\bibinfo {title} {Three-dimensional dirac semimetal and quantum transport in
  cd 3 as 2},}\ }\href@noop {} {\bibfield  {journal} {\bibinfo  {journal}
  {Phys. Rev. B}\ }\textbf {\bibinfo {volume} {88}},\ \bibinfo {pages} {125427}
  (\bibinfo {year} {2013})}\BibitemShut {NoStop}%
\bibitem [{\citenamefont {{Neupane}}\ \emph {et~al.}(2014)\citenamefont
  {{Neupane}}, \citenamefont {{Xu}}, \citenamefont {{Sankar}}, \citenamefont
  {{Alidoust}}, \citenamefont {{Bian}}, \citenamefont {{Liu}}, \citenamefont
  {{Belopolski}}, \citenamefont {{Chang}}, \citenamefont {{Jeng}},
  \citenamefont {{Lin}}, \citenamefont {{Bansil}}, \citenamefont {{Chou}},\
  and\ \citenamefont {{Hasan}}}]{neupane2014}%
  \BibitemOpen
  \bibfield  {author} {\bibinfo {author} {\bibfnamefont {M.}~\bibnamefont
  {{Neupane}}}, \bibinfo {author} {\bibfnamefont {S.-Y.}\ \bibnamefont {{Xu}}},
  \bibinfo {author} {\bibfnamefont {R.}~\bibnamefont {{Sankar}}}, \bibinfo
  {author} {\bibfnamefont {N.}~\bibnamefont {{Alidoust}}}, \bibinfo {author}
  {\bibfnamefont {G.}~\bibnamefont {{Bian}}}, \bibinfo {author} {\bibfnamefont
  {C.}~\bibnamefont {{Liu}}}, \bibinfo {author} {\bibfnamefont
  {I.}~\bibnamefont {{Belopolski}}}, \bibinfo {author} {\bibfnamefont {T.-R.}\
  \bibnamefont {{Chang}}}, \bibinfo {author} {\bibfnamefont {H.-T.}\
  \bibnamefont {{Jeng}}}, \bibinfo {author} {\bibfnamefont {H.}~\bibnamefont
  {{Lin}}}, \bibinfo {author} {\bibfnamefont {A.}~\bibnamefont {{Bansil}}},
  \bibinfo {author} {\bibfnamefont {F.}~\bibnamefont {{Chou}}}, \ and\ \bibinfo
  {author} {\bibfnamefont {M.~Z.}\ \bibnamefont {{Hasan}}},\ }\bibfield
  {title} {\enquote {\bibinfo {title} {{Observation of a three-dimensional
  topological Dirac semimetal phase in high-mobility Cd$_{3}$As$_{2}$}},}\
  }\href {\doibase 10.1038/ncomms4786} {\bibfield  {journal} {\bibinfo
  {journal} {Nat. Commun.}\ }\textbf {\bibinfo {volume} {5}},\ \bibinfo {eid}
  {3786} (\bibinfo {year} {2014})},\ \Eprint {http://arxiv.org/abs/1309.7892}
  {arXiv:1309.7892 [cond-mat.mes-hall]} \BibitemShut {NoStop}%
\bibitem [{\citenamefont {Xu}\ \emph {et~al.}(2015{\natexlab{b}})\citenamefont
  {Xu}, \citenamefont {Liu}, \citenamefont {Kushwaha}, \citenamefont {Sankar},
  \citenamefont {Krizan}, \citenamefont {Belopolski}, \citenamefont {Neupane},
  \citenamefont {Bian}, \citenamefont {Alidoust}, \citenamefont {Chang} \emph
  {et~al.}}]{xu2015observation}%
  \BibitemOpen
  \bibfield  {author} {\bibinfo {author} {\bibfnamefont {Su-Yang}\ \bibnamefont
  {Xu}}, \bibinfo {author} {\bibfnamefont {Chang}\ \bibnamefont {Liu}},
  \bibinfo {author} {\bibfnamefont {Satya~K}\ \bibnamefont {Kushwaha}},
  \bibinfo {author} {\bibfnamefont {Raman}\ \bibnamefont {Sankar}}, \bibinfo
  {author} {\bibfnamefont {Jason~W}\ \bibnamefont {Krizan}}, \bibinfo {author}
  {\bibfnamefont {Ilya}\ \bibnamefont {Belopolski}}, \bibinfo {author}
  {\bibfnamefont {Madhab}\ \bibnamefont {Neupane}}, \bibinfo {author}
  {\bibfnamefont {Guang}\ \bibnamefont {Bian}}, \bibinfo {author}
  {\bibfnamefont {Nasser}\ \bibnamefont {Alidoust}}, \bibinfo {author}
  {\bibfnamefont {Tay-Rong}\ \bibnamefont {Chang}},  \emph {et~al.},\
  }\bibfield  {title} {\enquote {\bibinfo {title} {Observation of fermi arc
  surface states in a topological metal},}\ }\href@noop {} {\bibfield
  {journal} {\bibinfo  {journal} {Science}\ }\textbf {\bibinfo {volume}
  {347}},\ \bibinfo {pages} {294--298} (\bibinfo {year}
  {2015}{\natexlab{b}})}\BibitemShut {NoStop}%
\bibitem [{\citenamefont {Liu}\ \emph {et~al.}(2014)\citenamefont {Liu},
  \citenamefont {Zhou}, \citenamefont {Zhang}, \citenamefont {Wang},
  \citenamefont {Weng}, \citenamefont {Prabhakaran}, \citenamefont {Mo},
  \citenamefont {Shen}, \citenamefont {Fang}, \citenamefont {Dai} \emph
  {et~al.}}]{liu2014discovery}%
  \BibitemOpen
  \bibfield  {author} {\bibinfo {author} {\bibfnamefont {ZK}~\bibnamefont
  {Liu}}, \bibinfo {author} {\bibfnamefont {B}~\bibnamefont {Zhou}}, \bibinfo
  {author} {\bibfnamefont {Y}~\bibnamefont {Zhang}}, \bibinfo {author}
  {\bibfnamefont {ZJ}~\bibnamefont {Wang}}, \bibinfo {author} {\bibfnamefont
  {HM}~\bibnamefont {Weng}}, \bibinfo {author} {\bibfnamefont {D}~\bibnamefont
  {Prabhakaran}}, \bibinfo {author} {\bibfnamefont {S-K}\ \bibnamefont {Mo}},
  \bibinfo {author} {\bibfnamefont {ZX}~\bibnamefont {Shen}}, \bibinfo {author}
  {\bibfnamefont {Z}~\bibnamefont {Fang}}, \bibinfo {author} {\bibfnamefont
  {X}~\bibnamefont {Dai}},  \emph {et~al.},\ }\bibfield  {title} {\enquote
  {\bibinfo {title} {Discovery of a three-dimensional topological dirac
  semimetal, na3bi},}\ }\href@noop {} {\bibfield  {journal} {\bibinfo
  {journal} {Science}\ }\textbf {\bibinfo {volume} {343}},\ \bibinfo {pages}
  {864--867} (\bibinfo {year} {2014})}\BibitemShut {NoStop}%
\bibitem [{\citenamefont {Borisenko}\ \emph {et~al.}(2014)\citenamefont
  {Borisenko}, \citenamefont {Gibson}, \citenamefont {Evtushinsky},
  \citenamefont {Zabolotnyy}, \citenamefont {B\"uchner},\ and\ \citenamefont
  {Cava}}]{Borisenko2014}%
  \BibitemOpen
  \bibfield  {author} {\bibinfo {author} {\bibfnamefont {Sergey}\ \bibnamefont
  {Borisenko}}, \bibinfo {author} {\bibfnamefont {Quinn}\ \bibnamefont
  {Gibson}}, \bibinfo {author} {\bibfnamefont {Danil}\ \bibnamefont
  {Evtushinsky}}, \bibinfo {author} {\bibfnamefont {Volodymyr}\ \bibnamefont
  {Zabolotnyy}}, \bibinfo {author} {\bibfnamefont {Bernd}\ \bibnamefont
  {B\"uchner}}, \ and\ \bibinfo {author} {\bibfnamefont {Robert~J.}\
  \bibnamefont {Cava}},\ }\bibfield  {title} {\enquote {\bibinfo {title}
  {Experimental realization of a three-dimensional dirac semimetal},}\ }\href
  {\doibase 10.1103/PhysRevLett.113.027603} {\bibfield  {journal} {\bibinfo
  {journal} {Phys. Rev. Lett.}\ }\textbf {\bibinfo {volume} {113}},\ \bibinfo
  {pages} {027603} (\bibinfo {year} {2014})}\BibitemShut {NoStop}%
\bibitem [{\citenamefont {Burkov}\ \emph {et~al.}(2011)\citenamefont {Burkov},
  \citenamefont {Hook},\ and\ \citenamefont {Balents}}]{Burkov2011nodal}%
  \BibitemOpen
  \bibfield  {author} {\bibinfo {author} {\bibfnamefont {A.~A.}\ \bibnamefont
  {Burkov}}, \bibinfo {author} {\bibfnamefont {M.~D.}\ \bibnamefont {Hook}}, \
  and\ \bibinfo {author} {\bibfnamefont {Leon}\ \bibnamefont {Balents}},\
  }\bibfield  {title} {\enquote {\bibinfo {title} {Topological nodal
  semimetals},}\ }\href {\doibase 10.1103/PhysRevB.84.235126} {\bibfield
  {journal} {\bibinfo  {journal} {Phys. Rev. B}\ }\textbf {\bibinfo {volume}
  {84}},\ \bibinfo {pages} {235126} (\bibinfo {year} {2011})}\BibitemShut
  {NoStop}%
\bibitem [{\citenamefont {{Bzdu{\v s}ek}}\ \emph {et~al.}(2016)\citenamefont
  {{Bzdu{\v s}ek}}, \citenamefont {{Wu}}, \citenamefont {{R{\"u}egg}},
  \citenamefont {{Sigrist}},\ and\ \citenamefont {{Soluyanov}}}]{Bzdusek2016}%
  \BibitemOpen
  \bibfield  {author} {\bibinfo {author} {\bibfnamefont {T.}~\bibnamefont
  {{Bzdu{\v s}ek}}}, \bibinfo {author} {\bibfnamefont {Q.}~\bibnamefont
  {{Wu}}}, \bibinfo {author} {\bibfnamefont {A.}~\bibnamefont {{R{\"u}egg}}},
  \bibinfo {author} {\bibfnamefont {M.}~\bibnamefont {{Sigrist}}}, \ and\
  \bibinfo {author} {\bibfnamefont {A.~A.}\ \bibnamefont {{Soluyanov}}},\
  }\bibfield  {title} {\enquote {\bibinfo {title} {{Nodal-chain metals}},}\
  }\href {\doibase 10.1038/nature19099} {\bibfield  {journal} {\bibinfo
  {journal} {Nature}\ }\textbf {\bibinfo {volume} {538}},\ \bibinfo {pages}
  {75--78} (\bibinfo {year} {2016})}\BibitemShut {NoStop}%
\bibitem [{\citenamefont {Chen}\ \emph {et~al.}(2017)\citenamefont {Chen},
  \citenamefont {Lu},\ and\ \citenamefont {Hou}}]{hopflink}%
  \BibitemOpen
  \bibfield  {author} {\bibinfo {author} {\bibfnamefont {Wei}\ \bibnamefont
  {Chen}}, \bibinfo {author} {\bibfnamefont {Hai-Zhou}\ \bibnamefont {Lu}}, \
  and\ \bibinfo {author} {\bibfnamefont {Jing-Min}\ \bibnamefont {Hou}},\
  }\bibfield  {title} {\enquote {\bibinfo {title} {Topological semimetals with
  a double-helix nodal link},}\ }\href {\doibase 10.1103/PhysRevB.96.041102}
  {\bibfield  {journal} {\bibinfo  {journal} {Phys. Rev. B}\ }\textbf {\bibinfo
  {volume} {96}},\ \bibinfo {pages} {041102} (\bibinfo {year}
  {2017})}\BibitemShut {NoStop}%
\bibitem [{\citenamefont {Yan}\ \emph {et~al.}(2017)\citenamefont {Yan},
  \citenamefont {Bi}, \citenamefont {Shen}, \citenamefont {Lu}, \citenamefont
  {Zhang},\ and\ \citenamefont {Wang}}]{nodal-link}%
  \BibitemOpen
  \bibfield  {author} {\bibinfo {author} {\bibfnamefont {Zhongbo}\ \bibnamefont
  {Yan}}, \bibinfo {author} {\bibfnamefont {Ren}\ \bibnamefont {Bi}}, \bibinfo
  {author} {\bibfnamefont {Huitao}\ \bibnamefont {Shen}}, \bibinfo {author}
  {\bibfnamefont {Ling}\ \bibnamefont {Lu}}, \bibinfo {author} {\bibfnamefont
  {Shou-Cheng}\ \bibnamefont {Zhang}}, \ and\ \bibinfo {author} {\bibfnamefont
  {Zhong}\ \bibnamefont {Wang}},\ }\bibfield  {title} {\enquote {\bibinfo
  {title} {Nodal-link semimetals},}\ }\href {\doibase
  10.1103/PhysRevB.96.041103} {\bibfield  {journal} {\bibinfo  {journal} {Phys.
  Rev. B}\ }\textbf {\bibinfo {volume} {96}},\ \bibinfo {pages} {041103}
  (\bibinfo {year} {2017})}\BibitemShut {NoStop}%
\bibitem [{\citenamefont {{Bi}}\ \emph {et~al.}(2017)\citenamefont {{Bi}},
  \citenamefont {{Yan}}, \citenamefont {{Lu}},\ and\ \citenamefont
  {{Wang}}}]{nodal-knot}%
  \BibitemOpen
  \bibfield  {author} {\bibinfo {author} {\bibfnamefont {R.}~\bibnamefont
  {{Bi}}}, \bibinfo {author} {\bibfnamefont {Z.}~\bibnamefont {{Yan}}},
  \bibinfo {author} {\bibfnamefont {L.}~\bibnamefont {{Lu}}}, \ and\ \bibinfo
  {author} {\bibfnamefont {Z.}~\bibnamefont {{Wang}}},\ }\bibfield  {title}
  {\enquote {\bibinfo {title} {{Nodal-knot semimetals}},}\ }\href@noop {}
  {\bibfield  {journal} {\bibinfo  {journal} {ArXiv e-prints}\ } (\bibinfo
  {year} {2017})},\ \Eprint {http://arxiv.org/abs/1704.06849} {arXiv:1704.06849
  [cond-mat.str-el]} \BibitemShut {NoStop}%
\bibitem [{\citenamefont {Nielsen}\ and\ \citenamefont
  {Ninomiya}(1981)}]{NIELSEN1981}%
  \BibitemOpen
  \bibfield  {author} {\bibinfo {author} {\bibfnamefont {H.~B.}\ \bibnamefont
  {Nielsen}}\ and\ \bibinfo {author} {\bibfnamefont {M.}~\bibnamefont
  {Ninomiya}},\ }\href@noop {} {\bibfield  {journal} {\bibinfo  {journal}
  {Nucl. Phys. B}\ }\textbf {\bibinfo {volume} {185}},\ \bibinfo {pages} {20}
  (\bibinfo {year} {1981})}\BibitemShut {NoStop}%
\bibitem [{\citenamefont {{Hosur}}\ and\ \citenamefont
  {{Qi}}(2013)}]{Hosur2013}%
  \BibitemOpen
  \bibfield  {author} {\bibinfo {author} {\bibfnamefont {P.}~\bibnamefont
  {{Hosur}}}\ and\ \bibinfo {author} {\bibfnamefont {X.}~\bibnamefont {{Qi}}},\
  }\bibfield  {title} {\enquote {\bibinfo {title} {{Recent developments in
  transport phenomena in Weyl semimetals}},}\ }\href {\doibase
  10.1016/j.crhy.2013.10.010} {\bibfield  {journal} {\bibinfo  {journal}
  {Comptes Rendus Physique}\ }\textbf {\bibinfo {volume} {14}},\ \bibinfo
  {pages} {857--870} (\bibinfo {year} {2013})},\ \Eprint
  {http://arxiv.org/abs/1309.4464} {arXiv:1309.4464 [cond-mat.str-el]}
  \BibitemShut {NoStop}%
\bibitem [{\citenamefont {Yan}\ and\ \citenamefont
  {Felser}(2017)}]{Yan2016review}%
  \BibitemOpen
  \bibfield  {author} {\bibinfo {author} {\bibfnamefont {Binghai}\ \bibnamefont
  {Yan}}\ and\ \bibinfo {author} {\bibfnamefont {Claudia}\ \bibnamefont
  {Felser}},\ }\bibfield  {title} {\enquote {\bibinfo {title} {Topological
  materials: Weyl semimetals},}\ }\href {\doibase
  10.1146/annurev-conmatphys-031016-025458} {\bibfield  {journal} {\bibinfo
  {journal} {Annual Review of Condensed Matter Physics}\ }\textbf {\bibinfo
  {volume} {8}},\ \bibinfo {pages} {337--354} (\bibinfo {year}
  {2017})}\BibitemShut {NoStop}%
\bibitem [{\citenamefont {Lu}\ and\ \citenamefont {Shen}(2016)}]{Lu2016review}%
  \BibitemOpen
  \bibfield  {author} {\bibinfo {author} {\bibfnamefont {Hai-Zhou}\
  \bibnamefont {Lu}}\ and\ \bibinfo {author} {\bibfnamefont {Shun-Qing}\
  \bibnamefont {Shen}},\ }\bibfield  {title} {\enquote {\bibinfo {title}
  {Quantum transport in topological semimetals under magnetic fields},}\ }\href
  {\doibase 10.1007/s11467-016-0609-y} {\bibfield  {journal} {\bibinfo
  {journal} {Frontiers of Physics}\ }\textbf {\bibinfo {volume} {12}},\
  \bibinfo {pages} {127201} (\bibinfo {year} {2016})}\BibitemShut {NoStop}%
\bibitem [{\citenamefont {Hasan}\ \emph {et~al.}(2017)\citenamefont {Hasan},
  \citenamefont {Xu}, \citenamefont {Belopolski},\ and\ \citenamefont
  {Huang}}]{Hasan2017review}%
  \BibitemOpen
  \bibfield  {author} {\bibinfo {author} {\bibfnamefont {M.~Zahid}\
  \bibnamefont {Hasan}}, \bibinfo {author} {\bibfnamefont {Su-Yang}\
  \bibnamefont {Xu}}, \bibinfo {author} {\bibfnamefont {Ilya}\ \bibnamefont
  {Belopolski}}, \ and\ \bibinfo {author} {\bibfnamefont {Shin-Ming}\
  \bibnamefont {Huang}},\ }\bibfield  {title} {\enquote {\bibinfo {title}
  {Discovery of weyl fermion semimetals and topological fermi arc states},}\
  }\href {\doibase 10.1146/annurev-conmatphys-031016-025225} {\bibfield
  {journal} {\bibinfo  {journal} {Annual Review of Condensed Matter Physics}\
  }\textbf {\bibinfo {volume} {8}},\ \bibinfo {pages} {289--309} (\bibinfo
  {year} {2017})}\BibitemShut {NoStop}%
\bibitem [{\citenamefont {{Burkov}}(2017)}]{Burkov2017review}%
  \BibitemOpen
  \bibfield  {author} {\bibinfo {author} {\bibfnamefont {A.~A.}\ \bibnamefont
  {{Burkov}}},\ }\bibfield  {title} {\enquote {\bibinfo {title} {{Weyl
  Metals}},}\ }\href@noop {} {\bibfield  {journal} {\bibinfo  {journal} {ArXiv
  e-prints}\ } (\bibinfo {year} {2017})},\ \Eprint
  {http://arxiv.org/abs/1704.06660} {arXiv:1704.06660 [cond-mat.mes-hall]}
  \BibitemShut {NoStop}%
\bibitem [{\citenamefont {Xu}\ \emph {et~al.}(2011)\citenamefont {Xu},
  \citenamefont {Weng}, \citenamefont {Wang}, \citenamefont {Dai},\ and\
  \citenamefont {Fang}}]{Xu2011doubleweyl}%
  \BibitemOpen
  \bibfield  {author} {\bibinfo {author} {\bibfnamefont {Gang}\ \bibnamefont
  {Xu}}, \bibinfo {author} {\bibfnamefont {Hongming}\ \bibnamefont {Weng}},
  \bibinfo {author} {\bibfnamefont {Zhijun}\ \bibnamefont {Wang}}, \bibinfo
  {author} {\bibfnamefont {Xi}~\bibnamefont {Dai}}, \ and\ \bibinfo {author}
  {\bibfnamefont {Zhong}\ \bibnamefont {Fang}},\ }\bibfield  {title} {\enquote
  {\bibinfo {title} {Chern semimetal and the quantized anomalous hall effect in
  ${\mathrm{hgcr}}_{2}{\mathrm{se}}_{4}$},}\ }\href {\doibase
  10.1103/PhysRevLett.107.186806} {\bibfield  {journal} {\bibinfo  {journal}
  {Phys. Rev. Lett.}\ }\textbf {\bibinfo {volume} {107}},\ \bibinfo {pages}
  {186806} (\bibinfo {year} {2011})}\BibitemShut {NoStop}%
\bibitem [{\citenamefont {Fang}\ \emph {et~al.}(2012)\citenamefont {Fang},
  \citenamefont {Gilbert}, \citenamefont {Dai},\ and\ \citenamefont
  {Bernevig}}]{Fang2012MW}%
  \BibitemOpen
  \bibfield  {author} {\bibinfo {author} {\bibfnamefont {Chen}\ \bibnamefont
  {Fang}}, \bibinfo {author} {\bibfnamefont {Matthew~J.}\ \bibnamefont
  {Gilbert}}, \bibinfo {author} {\bibfnamefont {Xi}~\bibnamefont {Dai}}, \ and\
  \bibinfo {author} {\bibfnamefont {B.~A.}\ \bibnamefont {Bernevig}},\
  }\bibfield  {title} {\enquote {\bibinfo {title} {Multi-weyl topological
  semimetals stabilized by point group symmetry},}\ }\href {\doibase
  10.1103/PhysRevLett.108.266802} {\bibfield  {journal} {\bibinfo  {journal}
  {Phys. Rev. Lett.}\ }\textbf {\bibinfo {volume} {108}},\ \bibinfo {pages}
  {266802} (\bibinfo {year} {2012})}\BibitemShut {NoStop}%
\bibitem [{\citenamefont {Huang}\ \emph {et~al.}(2016)\citenamefont {Huang},
  \citenamefont {Xu}, \citenamefont {Belopolski}, \citenamefont {Lee},
  \citenamefont {Chang}, \citenamefont {Chang}, \citenamefont {Wang},
  \citenamefont {Alidoust}, \citenamefont {Bian}, \citenamefont {Neupane},
  \citenamefont {Sanchez}, \citenamefont {Zheng}, \citenamefont {Jeng},
  \citenamefont {Bansil}, \citenamefont {Neupert}, \citenamefont {Lin},\ and\
  \citenamefont {Hasan}}]{Huang2016double}%
  \BibitemOpen
  \bibfield  {author} {\bibinfo {author} {\bibfnamefont {Shin-Ming}\
  \bibnamefont {Huang}}, \bibinfo {author} {\bibfnamefont {Su-Yang}\
  \bibnamefont {Xu}}, \bibinfo {author} {\bibfnamefont {Ilya}\ \bibnamefont
  {Belopolski}}, \bibinfo {author} {\bibfnamefont {Chi-Cheng}\ \bibnamefont
  {Lee}}, \bibinfo {author} {\bibfnamefont {Guoqing}\ \bibnamefont {Chang}},
  \bibinfo {author} {\bibfnamefont {Tay-Rong}\ \bibnamefont {Chang}}, \bibinfo
  {author} {\bibfnamefont {BaoKai}\ \bibnamefont {Wang}}, \bibinfo {author}
  {\bibfnamefont {Nasser}\ \bibnamefont {Alidoust}}, \bibinfo {author}
  {\bibfnamefont {Guang}\ \bibnamefont {Bian}}, \bibinfo {author}
  {\bibfnamefont {Madhab}\ \bibnamefont {Neupane}}, \bibinfo {author}
  {\bibfnamefont {Daniel}\ \bibnamefont {Sanchez}}, \bibinfo {author}
  {\bibfnamefont {Hao}\ \bibnamefont {Zheng}}, \bibinfo {author} {\bibfnamefont
  {Horng-Tay}\ \bibnamefont {Jeng}}, \bibinfo {author} {\bibfnamefont {Arun}\
  \bibnamefont {Bansil}}, \bibinfo {author} {\bibfnamefont {Titus}\
  \bibnamefont {Neupert}}, \bibinfo {author} {\bibfnamefont {Hsin}\
  \bibnamefont {Lin}}, \ and\ \bibinfo {author} {\bibfnamefont {M.~Zahid}\
  \bibnamefont {Hasan}},\ }\bibfield  {title} {\enquote {\bibinfo {title} {New
  type of weyl semimetal with quadratic double weyl fermions},}\ }\href
  {\doibase 10.1073/pnas.1514581113} {\bibfield  {journal} {\bibinfo  {journal}
  {Proceedings of the National Academy of Sciences}\ }\textbf {\bibinfo
  {volume} {113}},\ \bibinfo {pages} {1180--1185} (\bibinfo {year}
  {2016})}\BibitemShut {NoStop}%
\bibitem [{\citenamefont {Lai}(2015)}]{Lai2015dwsm}%
  \BibitemOpen
  \bibfield  {author} {\bibinfo {author} {\bibfnamefont {Hsin-Hua}\
  \bibnamefont {Lai}},\ }\bibfield  {title} {\enquote {\bibinfo {title}
  {Correlation effects in double-weyl semimetals},}\ }\href {\doibase
  10.1103/PhysRevB.91.235131} {\bibfield  {journal} {\bibinfo  {journal} {Phys.
  Rev. B}\ }\textbf {\bibinfo {volume} {91}},\ \bibinfo {pages} {235131}
  (\bibinfo {year} {2015})}\BibitemShut {NoStop}%
\bibitem [{\citenamefont {Jian}\ and\ \citenamefont
  {Yao}(2015)}]{Jian2015dwsm}%
  \BibitemOpen
  \bibfield  {author} {\bibinfo {author} {\bibfnamefont {Shao-Kai}\
  \bibnamefont {Jian}}\ and\ \bibinfo {author} {\bibfnamefont {Hong}\
  \bibnamefont {Yao}},\ }\bibfield  {title} {\enquote {\bibinfo {title}
  {Correlated double-weyl semimetals with coulomb interactions: Possible
  applications to ${\mathrm{hgcr}}_{2}{\mathrm{se}}_{4}$ and
  ${\mathrm{srsi}}_{2}$},}\ }\href {\doibase 10.1103/PhysRevB.92.045121}
  {\bibfield  {journal} {\bibinfo  {journal} {Phys. Rev. B}\ }\textbf {\bibinfo
  {volume} {92}},\ \bibinfo {pages} {045121} (\bibinfo {year}
  {2015})}\BibitemShut {NoStop}%
\bibitem [{\citenamefont {{Zhang}}\ \emph {et~al.}(2016)\citenamefont
  {{Zhang}}, \citenamefont {{Jian}},\ and\ \citenamefont
  {{Yao}}}]{Zhang2016triple}%
  \BibitemOpen
  \bibfield  {author} {\bibinfo {author} {\bibfnamefont {S.-X.}\ \bibnamefont
  {{Zhang}}}, \bibinfo {author} {\bibfnamefont {S.-K.}\ \bibnamefont {{Jian}}},
  \ and\ \bibinfo {author} {\bibfnamefont {H.}~\bibnamefont {{Yao}}},\
  }\bibfield  {title} {\enquote {\bibinfo {title} {{Correlated triple-Weyl
  semimetals with Coulomb interactions}},}\ }\href@noop {} {\bibfield
  {journal} {\bibinfo  {journal} {ArXiv e-prints}\ } (\bibinfo {year}
  {2016})},\ \Eprint {http://arxiv.org/abs/1610.08975} {arXiv:1610.08975
  [cond-mat.str-el]} \BibitemShut {NoStop}%
\bibitem [{\citenamefont {{Wang}}\ \emph {et~al.}(2016)\citenamefont {{Wang}},
  \citenamefont {{Liu}},\ and\ \citenamefont {{Zhang}}}]{Wang2016triple}%
  \BibitemOpen
  \bibfield  {author} {\bibinfo {author} {\bibfnamefont {J.-R.}\ \bibnamefont
  {{Wang}}}, \bibinfo {author} {\bibfnamefont {G.-Z.}\ \bibnamefont {{Liu}}}, \
  and\ \bibinfo {author} {\bibfnamefont {C.-J.}\ \bibnamefont {{Zhang}}},\
  }\bibfield  {title} {\enquote {\bibinfo {title} {{Anomalous violation of
  Fermi liquid theory in double- and triple-Weyl semimetals}},}\ }\href@noop {}
  {\bibfield  {journal} {\bibinfo  {journal} {ArXiv e-prints}\ } (\bibinfo
  {year} {2016})},\ \Eprint {http://arxiv.org/abs/1612.01729} {arXiv:1612.01729
  [cond-mat.str-el]} \BibitemShut {NoStop}%
\bibitem [{\citenamefont {Chen}\ and\ \citenamefont
  {Fiete}(2016)}]{Chen2016doubleweyl}%
  \BibitemOpen
  \bibfield  {author} {\bibinfo {author} {\bibfnamefont {Qi}~\bibnamefont
  {Chen}}\ and\ \bibinfo {author} {\bibfnamefont {Gregory~A.}\ \bibnamefont
  {Fiete}},\ }\bibfield  {title} {\enquote {\bibinfo {title} {Thermoelectric
  transport in double-weyl semimetals},}\ }\href {\doibase
  10.1103/PhysRevB.93.155125} {\bibfield  {journal} {\bibinfo  {journal} {Phys.
  Rev. B}\ }\textbf {\bibinfo {volume} {93}},\ \bibinfo {pages} {155125}
  (\bibinfo {year} {2016})}\BibitemShut {NoStop}%
\bibitem [{\citenamefont {Ahn}\ \emph {et~al.}(2016)\citenamefont {Ahn},
  \citenamefont {Hwang},\ and\ \citenamefont {Min}}]{Ahn2016multiWeyla}%
  \BibitemOpen
  \bibfield  {author} {\bibinfo {author} {\bibfnamefont {Seongjin}\
  \bibnamefont {Ahn}}, \bibinfo {author} {\bibfnamefont {E.~H.}\ \bibnamefont
  {Hwang}}, \ and\ \bibinfo {author} {\bibfnamefont {Hongki}\ \bibnamefont
  {Min}},\ }\bibfield  {title} {\enquote {\bibinfo {title} {Collective modes in
  multi-weyl semimetals},}\ }\href@noop {} {\bibfield  {journal} {\bibinfo
  {journal} {Scientific Reports}\ }\textbf {\bibinfo {volume} {6}},\ \bibinfo
  {pages} {34023} (\bibinfo {year} {2016})}\BibitemShut {NoStop}%
\bibitem [{\citenamefont {Ahn}\ \emph {et~al.}(2017)\citenamefont {Ahn},
  \citenamefont {Mele},\ and\ \citenamefont {Min}}]{Ahn2017mw}%
  \BibitemOpen
  \bibfield  {author} {\bibinfo {author} {\bibfnamefont {Seongjin}\
  \bibnamefont {Ahn}}, \bibinfo {author} {\bibfnamefont {E.~J.}\ \bibnamefont
  {Mele}}, \ and\ \bibinfo {author} {\bibfnamefont {Hongki}\ \bibnamefont
  {Min}},\ }\bibfield  {title} {\enquote {\bibinfo {title} {Optical
  conductivity of multi-weyl semimetals},}\ }\href {\doibase
  10.1103/PhysRevB.95.161112} {\bibfield  {journal} {\bibinfo  {journal} {Phys.
  Rev. B}\ }\textbf {\bibinfo {volume} {95}},\ \bibinfo {pages} {161112}
  (\bibinfo {year} {2017})}\BibitemShut {NoStop}%
\bibitem [{\citenamefont {{Hayata}}\ \emph {et~al.}(2017)\citenamefont
  {{Hayata}}, \citenamefont {{Kikuchi}},\ and\ \citenamefont
  {{Tanizaki}}}]{Hayata2017MW}%
  \BibitemOpen
  \bibfield  {author} {\bibinfo {author} {\bibfnamefont {T.}~\bibnamefont
  {{Hayata}}}, \bibinfo {author} {\bibfnamefont {Y.}~\bibnamefont {{Kikuchi}}},
  \ and\ \bibinfo {author} {\bibfnamefont {Y.}~\bibnamefont {{Tanizaki}}},\
  }\bibfield  {title} {\enquote {\bibinfo {title} {{Topological Properties of
  the Chiral Magnetic Effect in Multi-Weyl Semimetals}},}\ }\href@noop {}
  {\bibfield  {journal} {\bibinfo  {journal} {ArXiv e-prints}\ } (\bibinfo
  {year} {2017})},\ \Eprint {http://arxiv.org/abs/1703.02040} {arXiv:1703.02040
  [cond-mat.mes-hall]} \BibitemShut {NoStop}%
\bibitem [{\citenamefont {{Gupta}}(2017)}]{Gupta2017MW}%
  \BibitemOpen
  \bibfield  {author} {\bibinfo {author} {\bibfnamefont {A.}~\bibnamefont
  {{Gupta}}},\ }\bibfield  {title} {\enquote {\bibinfo {title} {{Floquet
  dynamics in multi-Weyl semimetals}},}\ }\href@noop {} {\bibfield  {journal}
  {\bibinfo  {journal} {ArXiv e-prints}\ } (\bibinfo {year} {2017})},\ \Eprint
  {http://arxiv.org/abs/1703.07271} {arXiv:1703.07271 [cond-mat.mes-hall]}
  \BibitemShut {NoStop}%
\bibitem [{\citenamefont {Roy}\ and\ \citenamefont
  {Sau}(2015)}]{roy_magnetic_2015}%
  \BibitemOpen
  \bibfield  {author} {\bibinfo {author} {\bibfnamefont {Bitan}\ \bibnamefont
  {Roy}}\ and\ \bibinfo {author} {\bibfnamefont {Jay~D.}\ \bibnamefont {Sau}},\
  }\bibfield  {title} {\enquote {\bibinfo {title} {Magnetic catalysis and
  axionic charge density wave in {{Weyl}} semimetals},}\ }\href {\doibase
  10.1103/PhysRevB.92.125141} {\bibfield  {journal} {\bibinfo  {journal} {Phys.
  Rev. B}\ }\textbf {\bibinfo {volume} {92}},\ \bibinfo {pages} {125141}
  (\bibinfo {year} {2015})}\BibitemShut {NoStop}%
\bibitem [{\citenamefont {Li}\ \emph {et~al.}(2016)\citenamefont {Li},
  \citenamefont {Roy},\ and\ \citenamefont {Das~Sarma}}]{li_weyl_2016}%
  \BibitemOpen
  \bibfield  {author} {\bibinfo {author} {\bibfnamefont {Xiao}\ \bibnamefont
  {Li}}, \bibinfo {author} {\bibfnamefont {Bitan}\ \bibnamefont {Roy}}, \ and\
  \bibinfo {author} {\bibfnamefont {S.}~\bibnamefont {Das~Sarma}},\ }\bibfield
  {title} {\enquote {\bibinfo {title} {Weyl fermions with arbitrary monopoles
  in magnetic fields: {{Landau}} levels, longitudinal magnetotransport, and
  density-wave ordering},}\ }\href {\doibase 10.1103/PhysRevB.94.195144}
  {\bibfield  {journal} {\bibinfo  {journal} {Phys. Rev. B}\ }\textbf {\bibinfo
  {volume} {94}},\ \bibinfo {pages} {195144} (\bibinfo {year}
  {2016})}\BibitemShut {NoStop}%
\bibitem [{\citenamefont {Yang}\ \emph {et~al.}(2011)\citenamefont {Yang},
  \citenamefont {Lu},\ and\ \citenamefont {Ran}}]{Yang2011}%
  \BibitemOpen
  \bibfield  {author} {\bibinfo {author} {\bibfnamefont {Kai-Yu}\ \bibnamefont
  {Yang}}, \bibinfo {author} {\bibfnamefont {Yuan-Ming}\ \bibnamefont {Lu}}, \
  and\ \bibinfo {author} {\bibfnamefont {Ying}\ \bibnamefont {Ran}},\
  }\bibfield  {title} {\enquote {\bibinfo {title} {Quantum hall effects in a
  weyl semimetal: Possible application in pyrochlore iridates},}\ }\href
  {\doibase 10.1103/PhysRevB.84.075129} {\bibfield  {journal} {\bibinfo
  {journal} {Phys. Rev. B}\ }\textbf {\bibinfo {volume} {84}},\ \bibinfo
  {pages} {075129} (\bibinfo {year} {2011})}\BibitemShut {NoStop}%
\bibitem [{\citenamefont {Burkov}(2014)}]{Burkov2014AHE}%
  \BibitemOpen
  \bibfield  {author} {\bibinfo {author} {\bibfnamefont {A.~A.}\ \bibnamefont
  {Burkov}},\ }\bibfield  {title} {\enquote {\bibinfo {title} {Anomalous hall
  effect in weyl metals},}\ }\href {\doibase 10.1103/PhysRevLett.113.187202}
  {\bibfield  {journal} {\bibinfo  {journal} {Phys. Rev. Lett.}\ }\textbf
  {\bibinfo {volume} {113}},\ \bibinfo {pages} {187202} (\bibinfo {year}
  {2014})}\BibitemShut {NoStop}%
\bibitem [{\citenamefont {Chan}\ \emph {et~al.}(2016)\citenamefont {Chan},
  \citenamefont {Lee}, \citenamefont {Burch}, \citenamefont {Han},\ and\
  \citenamefont {Ran}}]{Chan2016hall}%
  \BibitemOpen
  \bibfield  {author} {\bibinfo {author} {\bibfnamefont {Ching-Kit}\
  \bibnamefont {Chan}}, \bibinfo {author} {\bibfnamefont {Patrick~A.}\
  \bibnamefont {Lee}}, \bibinfo {author} {\bibfnamefont {Kenneth~S.}\
  \bibnamefont {Burch}}, \bibinfo {author} {\bibfnamefont {Jung~Hoon}\
  \bibnamefont {Han}}, \ and\ \bibinfo {author} {\bibfnamefont {Ying}\
  \bibnamefont {Ran}},\ }\bibfield  {title} {\enquote {\bibinfo {title} {When
  chiral photons meet chiral fermions: Photoinduced anomalous hall effects in
  weyl semimetals},}\ }\href {\doibase 10.1103/PhysRevLett.116.026805}
  {\bibfield  {journal} {\bibinfo  {journal} {Phys. Rev. Lett.}\ }\textbf
  {\bibinfo {volume} {116}},\ \bibinfo {pages} {026805} (\bibinfo {year}
  {2016})}\BibitemShut {NoStop}%
\bibitem [{\citenamefont {Yan}\ and\ \citenamefont
  {Wang}(2016)}]{Yan2016tunable}%
  \BibitemOpen
  \bibfield  {author} {\bibinfo {author} {\bibfnamefont {Zhongbo}\ \bibnamefont
  {Yan}}\ and\ \bibinfo {author} {\bibfnamefont {Zhong}\ \bibnamefont {Wang}},\
  }\bibfield  {title} {\enquote {\bibinfo {title} {Tunable weyl points in
  periodically driven nodal line semimetals},}\ }\href {\doibase
  10.1103/PhysRevLett.117.087402} {\bibfield  {journal} {\bibinfo  {journal}
  {Phys. Rev. Lett.}\ }\textbf {\bibinfo {volume} {117}},\ \bibinfo {pages}
  {087402} (\bibinfo {year} {2016})}\BibitemShut {NoStop}%
\bibitem [{\citenamefont {Son}\ and\ \citenamefont {Spivak}(2013)}]{son2012}%
  \BibitemOpen
  \bibfield  {author} {\bibinfo {author} {\bibfnamefont {D.~T.}\ \bibnamefont
  {Son}}\ and\ \bibinfo {author} {\bibfnamefont {B.~Z.}\ \bibnamefont
  {Spivak}},\ }\bibfield  {title} {\enquote {\bibinfo {title} {Chiral anomaly
  and classical negative magnetoresistance of weyl metals},}\ }\href {\doibase
  10.1103/PhysRevB.88.104412} {\bibfield  {journal} {\bibinfo  {journal} {Phys.
  Rev. B}\ }\textbf {\bibinfo {volume} {88}},\ \bibinfo {pages} {104412}
  (\bibinfo {year} {2013})}\BibitemShut {NoStop}%
\bibitem [{\citenamefont {Liu}\ \emph {et~al.}(2013)\citenamefont {Liu},
  \citenamefont {Ye},\ and\ \citenamefont {Qi}}]{liu2012}%
  \BibitemOpen
  \bibfield  {author} {\bibinfo {author} {\bibfnamefont {Chao-Xing}\
  \bibnamefont {Liu}}, \bibinfo {author} {\bibfnamefont {Peng}\ \bibnamefont
  {Ye}}, \ and\ \bibinfo {author} {\bibfnamefont {Xiao-Liang}\ \bibnamefont
  {Qi}},\ }\bibfield  {title} {\enquote {\bibinfo {title} {Chiral gauge field
  and axial anomaly in a weyl semimetal},}\ }\href {\doibase
  10.1103/PhysRevB.87.235306} {\bibfield  {journal} {\bibinfo  {journal} {Phys.
  Rev. B}\ }\textbf {\bibinfo {volume} {87}},\ \bibinfo {pages} {235306}
  (\bibinfo {year} {2013})}\BibitemShut {NoStop}%
\bibitem [{\citenamefont {Aji}(2012)}]{aji2011}%
  \BibitemOpen
  \bibfield  {author} {\bibinfo {author} {\bibfnamefont {Vivek}\ \bibnamefont
  {Aji}},\ }\bibfield  {title} {\enquote {\bibinfo {title} {Adler-bell-jackiw
  anomaly in weyl semimetals: Application to pyrochlore iridates},}\ }\href
  {\doibase 10.1103/PhysRevB.85.241101} {\bibfield  {journal} {\bibinfo
  {journal} {Phys. Rev. B}\ }\textbf {\bibinfo {volume} {85}},\ \bibinfo
  {pages} {241101} (\bibinfo {year} {2012})}\BibitemShut {NoStop}%
\bibitem [{\citenamefont {Zyuzin}\ and\ \citenamefont
  {Burkov}(2012)}]{Zyuzin2012anomaly}%
  \BibitemOpen
  \bibfield  {author} {\bibinfo {author} {\bibfnamefont {A.~A.}\ \bibnamefont
  {Zyuzin}}\ and\ \bibinfo {author} {\bibfnamefont {A.~A.}\ \bibnamefont
  {Burkov}},\ }\bibfield  {title} {\enquote {\bibinfo {title} {Topological
  response in weyl semimetals and the chiral anomaly},}\ }\href {\doibase
  10.1103/PhysRevB.86.115133} {\bibfield  {journal} {\bibinfo  {journal} {Phys.
  Rev. B}\ }\textbf {\bibinfo {volume} {86}},\ \bibinfo {pages} {115133}
  (\bibinfo {year} {2012})}\BibitemShut {NoStop}%
\bibitem [{\citenamefont {Wang}\ and\ \citenamefont {Zhang}(2013)}]{wang2013a}%
  \BibitemOpen
  \bibfield  {author} {\bibinfo {author} {\bibfnamefont {Zhong}\ \bibnamefont
  {Wang}}\ and\ \bibinfo {author} {\bibfnamefont {Shou-Cheng}\ \bibnamefont
  {Zhang}},\ }\bibfield  {title} {\enquote {\bibinfo {title} {Chiral anomaly,
  charge density waves, and axion strings from weyl semimetals},}\ }\href
  {\doibase 10.1103/PhysRevB.87.161107} {\bibfield  {journal} {\bibinfo
  {journal} {Phys. Rev. B}\ }\textbf {\bibinfo {volume} {87}},\ \bibinfo
  {pages} {161107} (\bibinfo {year} {2013})}\BibitemShut {NoStop}%
\bibitem [{\citenamefont {Hosur}\ and\ \citenamefont
  {Qi}(2015)}]{Hosur-anomaly}%
  \BibitemOpen
  \bibfield  {author} {\bibinfo {author} {\bibfnamefont {Pavan}\ \bibnamefont
  {Hosur}}\ and\ \bibinfo {author} {\bibfnamefont {Xiao-Liang}\ \bibnamefont
  {Qi}},\ }\bibfield  {title} {\enquote {\bibinfo {title} {Tunable circular
  dichroism due to the chiral anomaly in weyl semimetals},}\ }\href {\doibase
  10.1103/PhysRevB.91.081106} {\bibfield  {journal} {\bibinfo  {journal} {Phys.
  Rev. B}\ }\textbf {\bibinfo {volume} {91}},\ \bibinfo {pages} {081106}
  (\bibinfo {year} {2015})}\BibitemShut {NoStop}%
\bibitem [{\citenamefont {Kim}\ \emph {et~al.}(2013)\citenamefont {Kim},
  \citenamefont {Kim}, \citenamefont {Wang}, \citenamefont {Sasaki},
  \citenamefont {Satoh}, \citenamefont {Ohnishi}, \citenamefont {Kitaura},
  \citenamefont {Yang},\ and\ \citenamefont {Li}}]{Kim-chiral-anomaly}%
  \BibitemOpen
  \bibfield  {author} {\bibinfo {author} {\bibfnamefont {Heon-Jung}\
  \bibnamefont {Kim}}, \bibinfo {author} {\bibfnamefont {Ki-Seok}\ \bibnamefont
  {Kim}}, \bibinfo {author} {\bibfnamefont {J.-F.}\ \bibnamefont {Wang}},
  \bibinfo {author} {\bibfnamefont {M.}~\bibnamefont {Sasaki}}, \bibinfo
  {author} {\bibfnamefont {N.}~\bibnamefont {Satoh}}, \bibinfo {author}
  {\bibfnamefont {A.}~\bibnamefont {Ohnishi}}, \bibinfo {author} {\bibfnamefont
  {M.}~\bibnamefont {Kitaura}}, \bibinfo {author} {\bibfnamefont
  {M.}~\bibnamefont {Yang}}, \ and\ \bibinfo {author} {\bibfnamefont
  {L.}~\bibnamefont {Li}},\ }\bibfield  {title} {\enquote {\bibinfo {title}
  {Dirac versus weyl fermions in topological insulators: Adler-bell-jackiw
  anomaly in transport phenomena},}\ }\href {\doibase
  10.1103/PhysRevLett.111.246603} {\bibfield  {journal} {\bibinfo  {journal}
  {Phys. Rev. Lett.}\ }\textbf {\bibinfo {volume} {111}},\ \bibinfo {pages}
  {246603} (\bibinfo {year} {2013})}\BibitemShut {NoStop}%
\bibitem [{\citenamefont {Parameswaran}\ \emph {et~al.}(2014)\citenamefont
  {Parameswaran}, \citenamefont {Grover}, \citenamefont {Abanin}, \citenamefont
  {Pesin},\ and\ \citenamefont {Vishwanath}}]{Parameswaran-anomaly}%
  \BibitemOpen
  \bibfield  {author} {\bibinfo {author} {\bibfnamefont {S.~A.}\ \bibnamefont
  {Parameswaran}}, \bibinfo {author} {\bibfnamefont {T.}~\bibnamefont
  {Grover}}, \bibinfo {author} {\bibfnamefont {D.~A.}\ \bibnamefont {Abanin}},
  \bibinfo {author} {\bibfnamefont {D.~A.}\ \bibnamefont {Pesin}}, \ and\
  \bibinfo {author} {\bibfnamefont {A.}~\bibnamefont {Vishwanath}},\ }\bibfield
   {title} {\enquote {\bibinfo {title} {Probing the chiral anomaly with
  nonlocal transport in three-dimensional topological semimetals},}\ }\href
  {\doibase 10.1103/PhysRevX.4.031035} {\bibfield  {journal} {\bibinfo
  {journal} {Phys. Rev. X}\ }\textbf {\bibinfo {volume} {4}},\ \bibinfo {pages}
  {031035} (\bibinfo {year} {2014})}\BibitemShut {NoStop}%
\bibitem [{\citenamefont {Jiang}\ \emph {et~al.}(2007)\citenamefont {Jiang},
  \citenamefont {Henriksen}, \citenamefont {Tung}, \citenamefont {Wang},
  \citenamefont {Schwartz}, \citenamefont {Han}, \citenamefont {Kim},\ and\
  \citenamefont {Stormer}}]{Jiang2007moc}%
  \BibitemOpen
  \bibfield  {author} {\bibinfo {author} {\bibfnamefont {Z.}~\bibnamefont
  {Jiang}}, \bibinfo {author} {\bibfnamefont {E.~A.}\ \bibnamefont
  {Henriksen}}, \bibinfo {author} {\bibfnamefont {L.~C.}\ \bibnamefont {Tung}},
  \bibinfo {author} {\bibfnamefont {Y.-J.}\ \bibnamefont {Wang}}, \bibinfo
  {author} {\bibfnamefont {M.~E.}\ \bibnamefont {Schwartz}}, \bibinfo {author}
  {\bibfnamefont {M.~Y.}\ \bibnamefont {Han}}, \bibinfo {author} {\bibfnamefont
  {P.}~\bibnamefont {Kim}}, \ and\ \bibinfo {author} {\bibfnamefont {H.~L.}\
  \bibnamefont {Stormer}},\ }\bibfield  {title} {\enquote {\bibinfo {title}
  {Infrared spectroscopy of landau levels of graphene},}\ }\href {\doibase
  10.1103/PhysRevLett.98.197403} {\bibfield  {journal} {\bibinfo  {journal}
  {Phys. Rev. Lett.}\ }\textbf {\bibinfo {volume} {98}},\ \bibinfo {pages}
  {197403} (\bibinfo {year} {2007})}\BibitemShut {NoStop}%
\bibitem [{\citenamefont {Deacon}\ \emph {et~al.}(2007)\citenamefont {Deacon},
  \citenamefont {Chuang}, \citenamefont {Nicholas}, \citenamefont {Novoselov},\
  and\ \citenamefont {Geim}}]{Deacon2007moc}%
  \BibitemOpen
  \bibfield  {author} {\bibinfo {author} {\bibfnamefont {R.~S.}\ \bibnamefont
  {Deacon}}, \bibinfo {author} {\bibfnamefont {K.-C.}\ \bibnamefont {Chuang}},
  \bibinfo {author} {\bibfnamefont {R.~J.}\ \bibnamefont {Nicholas}}, \bibinfo
  {author} {\bibfnamefont {K.~S.}\ \bibnamefont {Novoselov}}, \ and\ \bibinfo
  {author} {\bibfnamefont {A.~K.}\ \bibnamefont {Geim}},\ }\bibfield  {title}
  {\enquote {\bibinfo {title} {Cyclotron resonance study of the electron and
  hole velocity in graphene monolayers},}\ }\href {\doibase
  10.1103/PhysRevB.76.081406} {\bibfield  {journal} {\bibinfo  {journal} {Phys.
  Rev. B}\ }\textbf {\bibinfo {volume} {76}},\ \bibinfo {pages} {081406}
  (\bibinfo {year} {2007})}\BibitemShut {NoStop}%
\bibitem [{\citenamefont {Plochocka}\ \emph {et~al.}(2008)\citenamefont
  {Plochocka}, \citenamefont {Faugeras}, \citenamefont {Orlita}, \citenamefont
  {Sadowski}, \citenamefont {Martinez}, \citenamefont {Potemski}, \citenamefont
  {Goerbig}, \citenamefont {Fuchs}, \citenamefont {Berger},\ and\ \citenamefont
  {de~Heer}}]{Plochocka2008moc}%
  \BibitemOpen
  \bibfield  {author} {\bibinfo {author} {\bibfnamefont {P.}~\bibnamefont
  {Plochocka}}, \bibinfo {author} {\bibfnamefont {C.}~\bibnamefont {Faugeras}},
  \bibinfo {author} {\bibfnamefont {M.}~\bibnamefont {Orlita}}, \bibinfo
  {author} {\bibfnamefont {M.~L.}\ \bibnamefont {Sadowski}}, \bibinfo {author}
  {\bibfnamefont {G.}~\bibnamefont {Martinez}}, \bibinfo {author}
  {\bibfnamefont {M.}~\bibnamefont {Potemski}}, \bibinfo {author}
  {\bibfnamefont {M.~O.}\ \bibnamefont {Goerbig}}, \bibinfo {author}
  {\bibfnamefont {J.-N.}\ \bibnamefont {Fuchs}}, \bibinfo {author}
  {\bibfnamefont {C.}~\bibnamefont {Berger}}, \ and\ \bibinfo {author}
  {\bibfnamefont {W.~A.}\ \bibnamefont {de~Heer}},\ }\bibfield  {title}
  {\enquote {\bibinfo {title} {High-energy limit of massless dirac fermions in
  multilayer graphene using magneto-optical transmission spectroscopy},}\
  }\href {\doibase 10.1103/PhysRevLett.100.087401} {\bibfield  {journal}
  {\bibinfo  {journal} {Phys. Rev. Lett.}\ }\textbf {\bibinfo {volume} {100}},\
  \bibinfo {pages} {087401} (\bibinfo {year} {2008})}\BibitemShut {NoStop}%
\bibitem [{\citenamefont {Schafgans}\ \emph {et~al.}(2012)\citenamefont
  {Schafgans}, \citenamefont {Post}, \citenamefont {Taskin}, \citenamefont
  {Ando}, \citenamefont {Qi}, \citenamefont {Chapler},\ and\ \citenamefont
  {Basov}}]{Schafgans2012moc}%
  \BibitemOpen
  \bibfield  {author} {\bibinfo {author} {\bibfnamefont {A.~A.}\ \bibnamefont
  {Schafgans}}, \bibinfo {author} {\bibfnamefont {K.~W.}\ \bibnamefont {Post}},
  \bibinfo {author} {\bibfnamefont {A.~A.}\ \bibnamefont {Taskin}}, \bibinfo
  {author} {\bibfnamefont {Yoichi}\ \bibnamefont {Ando}}, \bibinfo {author}
  {\bibfnamefont {Xiao-Liang}\ \bibnamefont {Qi}}, \bibinfo {author}
  {\bibfnamefont {B.~C.}\ \bibnamefont {Chapler}}, \ and\ \bibinfo {author}
  {\bibfnamefont {D.~N.}\ \bibnamefont {Basov}},\ }\bibfield  {title} {\enquote
  {\bibinfo {title} {Landau level spectroscopy of surface states in the
  topological insulator bi${}_{0.91}$sb${}_{0.09}$ via magneto-optics},}\
  }\href {\doibase 10.1103/PhysRevB.85.195440} {\bibfield  {journal} {\bibinfo
  {journal} {Phys. Rev. B}\ }\textbf {\bibinfo {volume} {85}},\ \bibinfo
  {pages} {195440} (\bibinfo {year} {2012})}\BibitemShut {NoStop}%
\bibitem [{\citenamefont {Orlita}\ \emph {et~al.}(2015)\citenamefont {Orlita},
  \citenamefont {Piot}, \citenamefont {Martinez}, \citenamefont {Kumar},
  \citenamefont {Faugeras}, \citenamefont {Potemski}, \citenamefont {Michel},
  \citenamefont {Hankiewicz}, \citenamefont {Brauner}, \citenamefont
  {Dra\ifmmode~\check{s}\else \v{s}\fi{}ar}, \citenamefont {Schreyeck},
  \citenamefont {Grauer}, \citenamefont {Brunner}, \citenamefont {Gould},
  \citenamefont {Br\"une},\ and\ \citenamefont {Molenkamp}}]{Orlita2015MOC}%
  \BibitemOpen
  \bibfield  {author} {\bibinfo {author} {\bibfnamefont {M.}~\bibnamefont
  {Orlita}}, \bibinfo {author} {\bibfnamefont {B.~A.}\ \bibnamefont {Piot}},
  \bibinfo {author} {\bibfnamefont {G.}~\bibnamefont {Martinez}}, \bibinfo
  {author} {\bibfnamefont {N.~K.~Sampath}\ \bibnamefont {Kumar}}, \bibinfo
  {author} {\bibfnamefont {C.}~\bibnamefont {Faugeras}}, \bibinfo {author}
  {\bibfnamefont {M.}~\bibnamefont {Potemski}}, \bibinfo {author}
  {\bibfnamefont {C.}~\bibnamefont {Michel}}, \bibinfo {author} {\bibfnamefont
  {E.~M.}\ \bibnamefont {Hankiewicz}}, \bibinfo {author} {\bibfnamefont
  {T.}~\bibnamefont {Brauner}}, \bibinfo {author} {\bibfnamefont {\ifmmode
  \check{C}\else~\v{C}\fi{}.}\ \bibnamefont {Dra\ifmmode~\check{s}\else
  \v{s}\fi{}ar}}, \bibinfo {author} {\bibfnamefont {S.}~\bibnamefont
  {Schreyeck}}, \bibinfo {author} {\bibfnamefont {S.}~\bibnamefont {Grauer}},
  \bibinfo {author} {\bibfnamefont {K.}~\bibnamefont {Brunner}}, \bibinfo
  {author} {\bibfnamefont {C.}~\bibnamefont {Gould}}, \bibinfo {author}
  {\bibfnamefont {C.}~\bibnamefont {Br\"une}}, \ and\ \bibinfo {author}
  {\bibfnamefont {L.~W.}\ \bibnamefont {Molenkamp}},\ }\bibfield  {title}
  {\enquote {\bibinfo {title} {Magneto-optics of massive dirac fermions in bulk
  ${\mathrm{bi}}_{2}{\mathrm{se}}_{3}$},}\ }\href {\doibase
  10.1103/PhysRevLett.114.186401} {\bibfield  {journal} {\bibinfo  {journal}
  {Phys. Rev. Lett.}\ }\textbf {\bibinfo {volume} {114}},\ \bibinfo {pages}
  {186401} (\bibinfo {year} {2015})}\BibitemShut {NoStop}%
\bibitem [{\citenamefont {Chen}\ \emph {et~al.}(2015)\citenamefont {Chen},
  \citenamefont {Chen}, \citenamefont {Song}, \citenamefont {Schneeloch},
  \citenamefont {Gu}, \citenamefont {Wang},\ and\ \citenamefont
  {Wang}}]{Chen2016ZrTe5}%
  \BibitemOpen
  \bibfield  {author} {\bibinfo {author} {\bibfnamefont {R.~Y.}\ \bibnamefont
  {Chen}}, \bibinfo {author} {\bibfnamefont {Z.~G.}\ \bibnamefont {Chen}},
  \bibinfo {author} {\bibfnamefont {X.-Y.}\ \bibnamefont {Song}}, \bibinfo
  {author} {\bibfnamefont {J.~A.}\ \bibnamefont {Schneeloch}}, \bibinfo
  {author} {\bibfnamefont {G.~D.}\ \bibnamefont {Gu}}, \bibinfo {author}
  {\bibfnamefont {F.}~\bibnamefont {Wang}}, \ and\ \bibinfo {author}
  {\bibfnamefont {N.~L.}\ \bibnamefont {Wang}},\ }\bibfield  {title} {\enquote
  {\bibinfo {title} {Magnetoinfrared spectroscopy of landau levels and zeeman
  splitting of three-dimensional massless dirac fermions in
  ${\mathrm{zrte}}_{5}$},}\ }\href {\doibase 10.1103/PhysRevLett.115.176404}
  {\bibfield  {journal} {\bibinfo  {journal} {Phys. Rev. Lett.}\ }\textbf
  {\bibinfo {volume} {115}},\ \bibinfo {pages} {176404} (\bibinfo {year}
  {2015})}\BibitemShut {NoStop}%
\bibitem [{\citenamefont {Akrap}\ \emph {et~al.}(2016)\citenamefont {Akrap},
  \citenamefont {Hakl}, \citenamefont {Tchoumakov}, \citenamefont {Crassee},
  \citenamefont {Kuba}, \citenamefont {Goerbig}, \citenamefont {Homes},
  \citenamefont {Caha}, \citenamefont {Nov\'ak}, \citenamefont {Teppe},
  \citenamefont {Desrat}, \citenamefont {Koohpayeh}, \citenamefont {Wu},
  \citenamefont {Armitage}, \citenamefont {Nateprov}, \citenamefont
  {Arushanov}, \citenamefont {Gibson}, \citenamefont {Cava}, \citenamefont
  {van~der Marel}, \citenamefont {Piot}, \citenamefont {Faugeras},
  \citenamefont {Martinez}, \citenamefont {Potemski},\ and\ \citenamefont
  {Orlita}}]{Akrap2016MOC}%
  \BibitemOpen
  \bibfield  {author} {\bibinfo {author} {\bibfnamefont {A.}~\bibnamefont
  {Akrap}}, \bibinfo {author} {\bibfnamefont {M.}~\bibnamefont {Hakl}},
  \bibinfo {author} {\bibfnamefont {S.}~\bibnamefont {Tchoumakov}}, \bibinfo
  {author} {\bibfnamefont {I.}~\bibnamefont {Crassee}}, \bibinfo {author}
  {\bibfnamefont {J.}~\bibnamefont {Kuba}}, \bibinfo {author} {\bibfnamefont
  {M.~O.}\ \bibnamefont {Goerbig}}, \bibinfo {author} {\bibfnamefont {C.~C.}\
  \bibnamefont {Homes}}, \bibinfo {author} {\bibfnamefont {O.}~\bibnamefont
  {Caha}}, \bibinfo {author} {\bibfnamefont {J.}~\bibnamefont {Nov\'ak}},
  \bibinfo {author} {\bibfnamefont {F.}~\bibnamefont {Teppe}}, \bibinfo
  {author} {\bibfnamefont {W.}~\bibnamefont {Desrat}}, \bibinfo {author}
  {\bibfnamefont {S.}~\bibnamefont {Koohpayeh}}, \bibinfo {author}
  {\bibfnamefont {L.}~\bibnamefont {Wu}}, \bibinfo {author} {\bibfnamefont
  {N.~P.}\ \bibnamefont {Armitage}}, \bibinfo {author} {\bibfnamefont
  {A.}~\bibnamefont {Nateprov}}, \bibinfo {author} {\bibfnamefont
  {E.}~\bibnamefont {Arushanov}}, \bibinfo {author} {\bibfnamefont {Q.~D.}\
  \bibnamefont {Gibson}}, \bibinfo {author} {\bibfnamefont {R.~J.}\
  \bibnamefont {Cava}}, \bibinfo {author} {\bibfnamefont {D.}~\bibnamefont
  {van~der Marel}}, \bibinfo {author} {\bibfnamefont {B.~A.}\ \bibnamefont
  {Piot}}, \bibinfo {author} {\bibfnamefont {C.}~\bibnamefont {Faugeras}},
  \bibinfo {author} {\bibfnamefont {G.}~\bibnamefont {Martinez}}, \bibinfo
  {author} {\bibfnamefont {M.}~\bibnamefont {Potemski}}, \ and\ \bibinfo
  {author} {\bibfnamefont {M.}~\bibnamefont {Orlita}},\ }\bibfield  {title}
  {\enquote {\bibinfo {title} {Magneto-optical signature of massless kane
  electrons in ${\mathrm{cd}}_{3}{\mathrm{as}}_{2}$},}\ }\href {\doibase
  10.1103/PhysRevLett.117.136401} {\bibfield  {journal} {\bibinfo  {journal}
  {Phys. Rev. Lett.}\ }\textbf {\bibinfo {volume} {117}},\ \bibinfo {pages}
  {136401} (\bibinfo {year} {2016})}\BibitemShut {NoStop}%
\bibitem [{\citenamefont {Gusynin}\ \emph {et~al.}(2007)\citenamefont
  {Gusynin}, \citenamefont {Sharapov},\ and\ \citenamefont
  {Carbotte}}]{Gusynin2007MOC}%
  \BibitemOpen
  \bibfield  {author} {\bibinfo {author} {\bibfnamefont {V.~P.}\ \bibnamefont
  {Gusynin}}, \bibinfo {author} {\bibfnamefont {S.~G.}\ \bibnamefont
  {Sharapov}}, \ and\ \bibinfo {author} {\bibfnamefont {J.~P.}\ \bibnamefont
  {Carbotte}},\ }\bibfield  {title} {\enquote {\bibinfo {title} {Anomalous
  absorption line in the magneto-optical response of graphene},}\ }\href
  {\doibase 10.1103/PhysRevLett.98.157402} {\bibfield  {journal} {\bibinfo
  {journal} {Phys. Rev. Lett.}\ }\textbf {\bibinfo {volume} {98}},\ \bibinfo
  {pages} {157402} (\bibinfo {year} {2007})}\BibitemShut {NoStop}%
\bibitem [{\citenamefont {Ashby}\ and\ \citenamefont
  {Carbotte}(2013)}]{Ashby2013weyl}%
  \BibitemOpen
  \bibfield  {author} {\bibinfo {author} {\bibfnamefont {Phillip E.~C.}\
  \bibnamefont {Ashby}}\ and\ \bibinfo {author} {\bibfnamefont {J.~P.}\
  \bibnamefont {Carbotte}},\ }\bibfield  {title} {\enquote {\bibinfo {title}
  {Magneto-optical conductivity of weyl semimetals},}\ }\href {\doibase
  10.1103/PhysRevB.87.245131} {\bibfield  {journal} {\bibinfo  {journal} {Phys.
  Rev. B}\ }\textbf {\bibinfo {volume} {87}},\ \bibinfo {pages} {245131}
  (\bibinfo {year} {2013})}\BibitemShut {NoStop}%
\bibitem [{\citenamefont {Malcolm}\ and\ \citenamefont
  {Nicol}(2014)}]{Malcolm2014moc}%
  \BibitemOpen
  \bibfield  {author} {\bibinfo {author} {\bibfnamefont {J.~D.}\ \bibnamefont
  {Malcolm}}\ and\ \bibinfo {author} {\bibfnamefont {E.~J.}\ \bibnamefont
  {Nicol}},\ }\bibfield  {title} {\enquote {\bibinfo {title} {Magneto-optics of
  general pseudospin-$s$ two-dimensional dirac-weyl fermions},}\ }\href
  {\doibase 10.1103/PhysRevB.90.035405} {\bibfield  {journal} {\bibinfo
  {journal} {Phys. Rev. B}\ }\textbf {\bibinfo {volume} {90}},\ \bibinfo
  {pages} {035405} (\bibinfo {year} {2014})}\BibitemShut {NoStop}%
\bibitem [{\citenamefont {Baltzer}\ \emph {et~al.}(1965)\citenamefont
  {Baltzer}, \citenamefont {Lehmann},\ and\ \citenamefont
  {Robbins}}]{baltzer1965}%
  \BibitemOpen
  \bibfield  {author} {\bibinfo {author} {\bibfnamefont {P.~K.}\ \bibnamefont
  {Baltzer}}, \bibinfo {author} {\bibfnamefont {H.~W.}\ \bibnamefont
  {Lehmann}}, \ and\ \bibinfo {author} {\bibfnamefont {M.}~\bibnamefont
  {Robbins}},\ }\bibfield  {title} {\enquote {\bibinfo {title} {Insulating
  {{Ferromagnetic Spinels}}},}\ }\href {\doibase 10.1103/PhysRevLett.15.493}
  {\bibfield  {journal} {\bibinfo  {journal} {Phys. Rev. Lett.}\ }\textbf
  {\bibinfo {volume} {15}},\ \bibinfo {pages} {493--495} (\bibinfo {year}
  {1965})}\BibitemShut {NoStop}%
\end{thebibliography}%

\end{document}